# First-principles-based Prediction of Phase Fields: Part II. Solid Solution Tunability in Refractory Complex Concentrated Alloys

Pravan Omprakash[1], Nicholas Crnkovich[2], Nathan Curtis[2], John Cavin[3], Adrien Couet[2], Rohan Mishra[3,1]

[1]Institute of Materials Science & Engineering, Washington University in St. Louis, St. Louis, MO, USA

[2]Department of Nuclear Engineering & Engineering Physics, University of Wisconsin, Madison, WI, USA

[3]Department of Mechanical Engineering & Materials Science, Washington University in St. Louis, St. Louis, MO, USA

Corresponding authors: P.O. (o.pravan@wustl.edu); R.M. (rmishra@wustl.edu)

## Abstract

Refractory complex concentrated alloys (RCCAs) offer a high-dimensional composition space for designing either robust single-phase solid solutions or controlled multiphase microstructures. However, most RCCA studies have examined isolated compositions, often near the equimolar point, rather than determining how solid-solution stability varies across the surrounding composition space. Here, we use rapid thermodynamic phase-field calculations to map the topology of solid-solution stability in RCCAs composed of Cr, Hf, Mo, Nb, Ta, Ti, V, W, and Zr. Benchmarking the model against 109 experimentally reported quaternary-to-senary compositions produced through diverse processing routes yields an overall classification accuracy of 80%. We then introduce percentage miscible region, a metric that separates robust solid-solution-forming systems from those exhibiting compositionally tunable microstructures. Analysis of ternary, quaternary, and quinary RCCAs reveals that higher-order miscibility is strongly inherited from constituent binary subsystems, indicating the importance of choosing suitable binary interactions when designing RCCAs. Element-specific trends further show that Ti, Ta, Nb, and Mo generally stabilize solid solutions, whereas Cr and Zr promote ordered precipitation and phase segregation. Hf, W, and V, meanwhile, enable tuning between these regimes. These results establish a composition-space framework for the targeted design of RCCAs with robust solid-solution stability or tunable multiphase microstructures.

## 1. Introduction

Multiple principal element alloys (MPEAs) broaden alloy design by enabling the exploration of large composition spaces [1]. In refractory complex concentrated alloys (RCCAs), this flexibility is especially valuable because several high-melting-point elements can be combined over broad compositional ranges to tune phase stability, microstructure, and high-temperature functionality [2] [3]. As a recent example, Zhang et al. [4] synthesized a RCCA with a coherent triphasic nanoarchitecture in which BCC, FCC, and HCP phases coexist through strain-induced coherency. Other studies have linked RCCA behavior to phase and defect structures, including kink-band-mediated fracture resistance in NbTaTiHf [5] and radiation-resistant W-based refractory high-entropy alloys designed by combining CALPHAD, cluster expansion and Monte Carlo simulations [6], [7]. The design of RCCAs involves navigating high-dimensional composition spaces in which different regions may stabilize single-phase solid solutions or multiphase segregated states including ordered precipitates. A nominally stable equimolar alloy could be surrounded by unstable regions that enter a miscibility gap or precipitate ordered phases. Conversely, an alloy system may contain a broad, continuous single-phase solid solution (SPSS) region that is tolerant to local compositional variations. Between these two extremes are compositionally sensitive systems, where small changes in alloy chemistry can be used to deliberately tune phase segregation or drive ordered precipitate formation. A high-dimensional phase field can contain a large SPSS basin or isolated miscible pockets, along with lower-order compositional regions where ordered precipitates become favorable. Mapping this topology distinguishes alloys that are robust to compositional fluctuations from compositionally tunable systems in which small changes in composition produce substantial changes in phase fractions and properties. This composition-space perspective is especially important during processing and operation, when local composition and microstructure can evolve through chemical fluctuations, segregation, diffusion, oxidation, or irradiation-induced transmutation. Thus, understanding the size, connectivity, and morphology of miscible regions is central to designing RCCAs with either robust solid-solution stability or controllable multiphase microstructures.

The unique topology of miscibility gaps in multicomponent alloys was highlighted by Morral et al. [10], who demonstrated that these gaps could adopt complex configurations such as isolated domains and "rose geometries" — that are not easily interpreted using conventional binary or ternary phase-diagram intuition. In a later study, Morral and Chen [11] extended this discussion from miscibility-gap equilibria to spinodal stability in high-entropy alloys. They established the conditions required for three-phase miscibility gaps and showed that spinodal surfaces are not confined to binary edges but can propagate deep into the higher-dimensional compositional space. While these foundational studies establish miscibility-gap topology as a central descriptor of multicomponent phase behavior, they relied on symmetric interaction parameters and neglected competing ordered intermetallics. The phase stability of alloys is rarely governed by a single solid solution alone; competing ordered intermetallics can severely restrict and reshape the accessible SPSS regions. Furthermore, while previous computational studies have successfully predicted the stability of specific, often equimolar, MPEAs [7–9], such targeted approaches do not capture how phase stability and miscibility-gap topologies continuously evolve across the broader multidimensional composition space.

In this Article, we use rapid thermodynamic phase-field calculations as a framework for analyzing the topology of refractory RCCA composition spaces. Using the model introduced in Part I accompanying this article — that combines first-principles binary mixing enthalpies with a sub-regular solution model — we examine quaternary and higher-order RCCAs composed of the nine refractory elements: Cr, Hf, Mo, Nb, Ta, Ti, V, W, and Zr. The model is first benchmarked against a diverse experimental dataset of 109 quaternary-to-senary compositions obtained using varied processing routes, yielding an overall classification accuracy of 80%. With this reliability established, we use the predicted phase fields to map how SPSS stability is distributed, connected, and chemically controlled across multicomponent composition spaces. To quantify this behavior, we introduce a metric termed as the *percentage miscible region* (PMR). High-PMR systems contain broad SPSS regions, low-PMR systems are dominated by phase segregation or ordered precipitation, and mid-PMR systems offer compositionally tunable transitions where modest alloy adjustments trigger precipitation or phase segregation. Analyzing RCCA composition spaces reveals that higher-order miscibility is strongly inherited from constituent binary and ternary subsystems, indicating that configurational entropy cannot overcome strong chemical interactions. Across the nine-element refractory space, Ti, Ta, Nb, and Mo generally stabilize SPSS regions, whereas Cr and Zr tend to promote phase segregation and precipitation. Hf, W and V can serve as effective tuning elements for mid-range PMR alloys. Ultimately, this computationally efficient framework establishes concrete design rules for rapidly mapping multinary phase fields and accelerating the development of RCCAs with targeted microstructures.

The remainder of this Article is organized as follows. In Section 2.1, we compare the model predictions with experimentally reported RCCAs and use SymPlex maps to illustrate the evolution of phase-fields with composition in quaternary and quinary systems. In Section 2.2, we introduce percentage miscible region (PMR) as a parameter to quantify the extent of SPSS stability across multicomponent composition spaces. In Section 2.3, we analyze how miscibility propagates from lower-order binary and ternary subsystems into higher-order phase fields and introduce rules to design RCCAs with large single phase solid solution regions. In Section 2.4, we examine the chemical origin of the solid-solution stability and identify elements that act as stabilizers or destabilizers of refractory SPSS phases. Finally, in Section 2.5, we investigate the role of temperature in simplifying phase equilibria and expanding the SPSS region in RCCAs.

## 2. Results and Discussion

### 2.1. Phase stability in selected experimental RHEAs

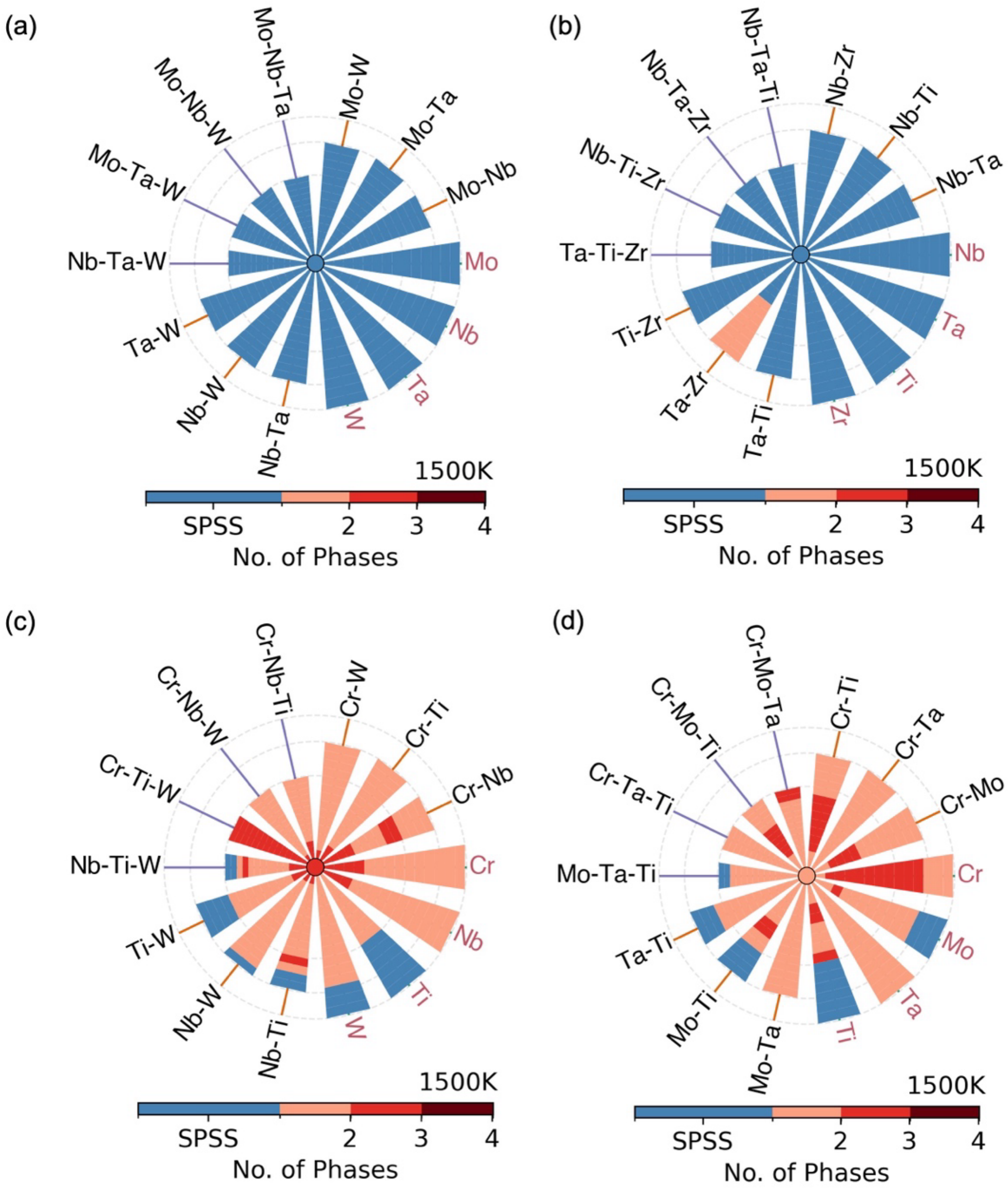

**Fig. 1. SymPlex maps of representative experimentally reported quaternary refractory alloy systems.** (a) Mo–Nb–Ta–W, (b) Nb–Ta–Ti–Zr, (c) Cr–Nb–Ti–W, and (d) Cr–Mo–Ta–Ti. The central node corresponds to the equimolar quaternary composition, while radial paths connect the equimolar point to unary, binary, and ternary equimolar endpoints. Blue regions indicate stable single-phase solid solution, while red regions indicate the number of phases in equilibrium.

Having benchmarked our model against binary and ternary systems in Part I, we now extend the analysis to experimental reports of quaternary, quinary, and senary RCCAs. By surveying the literature, and including databases by Gorsse et al. [8], Wei et al. [9] and Mohammadnejad et al. [10], we have compiled a dataset of 50 RCCAs containing the nine elements: Cr, Hf, Mo, Nb, Ta, Ti, V, W, and Zr., as summarized in Table 1 in Appendix A. To move beyond equimolar compositions, we include reported off-equimolar alloys and account for different processing routes for identical compositions, resulting in a total of 109 experimental data points. For annealed, sintered, or cold-worked samples, the reported annealing temperature ($T_{anneal}$) was used as the processing temperature ($T_{proc}$) for comparison with our model. For as-cast and additively manufactured samples, $T_{proc}$ was approximated as the lowest melting temperature of the constituent elements ($T_{melt}$) [11]. Because the as-cast solidification temperature depends on dendritic and interdendritic segregation, which lowers the temperature at which the last interdendritic liquid solidifies below the average value, this lowest melting temperature is used as a conservative proxy for the solidus temperature. Our workflow predicts the miscibility temperature ($T_{misc}$) for each composition; phase segregation is predicted if $T_{misc} > T_{proc}$, whereas an SPSS is expected if $T_{misc} \leq T_{proc}$.

Overall, the model correctly classifies 87 of the 109 compositions, yielding an 80% accuracy. The primary discrepancies occur in as-cast samples, mainly due to kinetic considerations, where the model predicts SPSS for eight samples that are experimentally phase segregated; conversely, five are predicted segregated but observed as SPSS. Additionally, five annealed and four AM samples were misclassified due to the non-equilibrium nature of synthesis, listed in detail in Appendix A. Taken separately, the 29 annealed samples, whose $T_{proc}$ is an experimentally reported temperature and therefore independent of this proxy, are classified with 83% accuracy (24 of 29). The complete benchmarking analysis, including the full comparison table, confusion matrices, and discussion of correctly classified and misclassified compositions, are provided in Appendix A.

In the following, we use the experimental dataset to identify alloy systems that capture the dominant phase-field behaviors observed across RCCAs. We focus on systems where the predicted phase fields, for quaternary systems and for quinary systems, illustrate the evolution of solid-solution stability across the composition space. We use SymPlex maps to conveniently visualize the high-dimensional space; these maps provide a two-dimensional projection of a multicomponent composition space along specific composition paths [12]. We choose pathways that connect lower-order equimolar compositions to the higher-order equimolar composition because they are intuitive for understanding alloy synthesis especially using additive manufacturing processes. Equimolar compositions are also representative of the combination of elements that either increase/decrease in mole fraction as we go along a path. In each map, the central node represents the equimolar quaternary or quinary composition, and each radial path connects this point to a lower-order equimolar endpoint: unary vertices, binary edges, ternary faces and so on. Moving outward along a path therefore corresponds to continuously enriching the alloy toward the composition indicated by the outer label. Blue regions indicate SPSS compositions, while regions in red represent phase segregated regions, demarcating the number of the phases in equilibrium. Thus, the maps reveal whether solid-solution stability is broadly connected across the phase space or localized to particular elemental combinations.

***2.1.1. Quaternary systems***. Fig. 1 presents four representative quaternary systems at 1500 K. Among these, the Mo–Nb–Ta–W system, shown in Fig. 1(a), serves as a robust SPSS-forming reference. The predicted solid-solution phase fraction remains close to unity across nearly the entire quaternary phase space, consistent with an experimental report of BCC solid-solution behavior in this alloy system [13]. The broad stability of this system indicates that Mo, Nb, Ta, and W form a mutually compatible BCC refractory alloy space, where local compositional fluctuations are unlikely to drive phase segregation at this temperature.

The Nb–Ta–Ti–Zr system, shown in Fig. 1(b), also retains a large, connected SPSS region, but its stability is less uniform than that of Mo–Nb–Ta–W. While the solid solution surrounding the equimolar quaternary composition remains largely stable, reduced solid-solution phase fractions appear when moving towards the Ta-Zr binary system, i.e., reducing the fraction of Mo and Nb. Out of six experimental reports on NbTaTiZr, we correctly predict four of them to be SPSS, as seen in Table 1. For two as-cast samples synthesized by Zyka et al. [14], one being Zr-rich and the other Nb-rich, segregation into multiple BCC phases is experimentally observed; we predict this segregation at 1000 K for the Zr-rich sample, whereas the Nb-rich sample is predicted to segregate only at or below about 700 K (Appendix D, Fig. 10(a,b)). The predictions for experimentally reported samples are present in the public repository along with the code to predict RCCA phase fields [15]. Additionally, a publicly accessible web-based interface has been developed to facilitate interactive exploration of the predicted phase diagrams, accessible at https://raptor.engr.wustl.edu.

The two Cr-containing quaternary systems, Cr–Nb–Ti–W and Cr–Mo–Ta–Ti, shown in Fig. 1(c-d), respectively, show a more pronounced loss of solid-solution stability. In Cr–Nb–Ti–W experimental observations, wherein as-cast alloys retain a BCC solid solution, annealed samples at 1673 K show phase segregation. The SymPlex map indicates that Cr-associated instability is uniform across the entire quaternary space. For the CrNbTiW phase at 1673 K, Coury et al. [16] predicted a dual BCC phase segregation, with one of the BCC phases enriched in Cr. We too predict two BCC phases, rich and lean in Cr with 26% and 74% phase fractions, respectively (Appendix D, Fig. 10(c)).

A more pronounced instability is observed in Cr–Mo–Ta–Ti, as shown in Fig. 1(d). The Mo–Ta–Ti-rich side retains a high solid-solution phase fraction, whereas Cr-rich paths and several Cr-containing subsystems show three-phase segregations. Experimentally, this system has been reported as a BCC solid solution in the as-cast condition that, upon annealing at 1673 K, forms a BCC solid solution with Laves precipitates [16]. Both of these conditions are correctly predicted by the model, as seen from Table 1. The predicted topology therefore successfully captures both the stable BCC solid solution regions promoted by Ti and the proximate Cr-containing regions that undergo secondary phase precipitation.

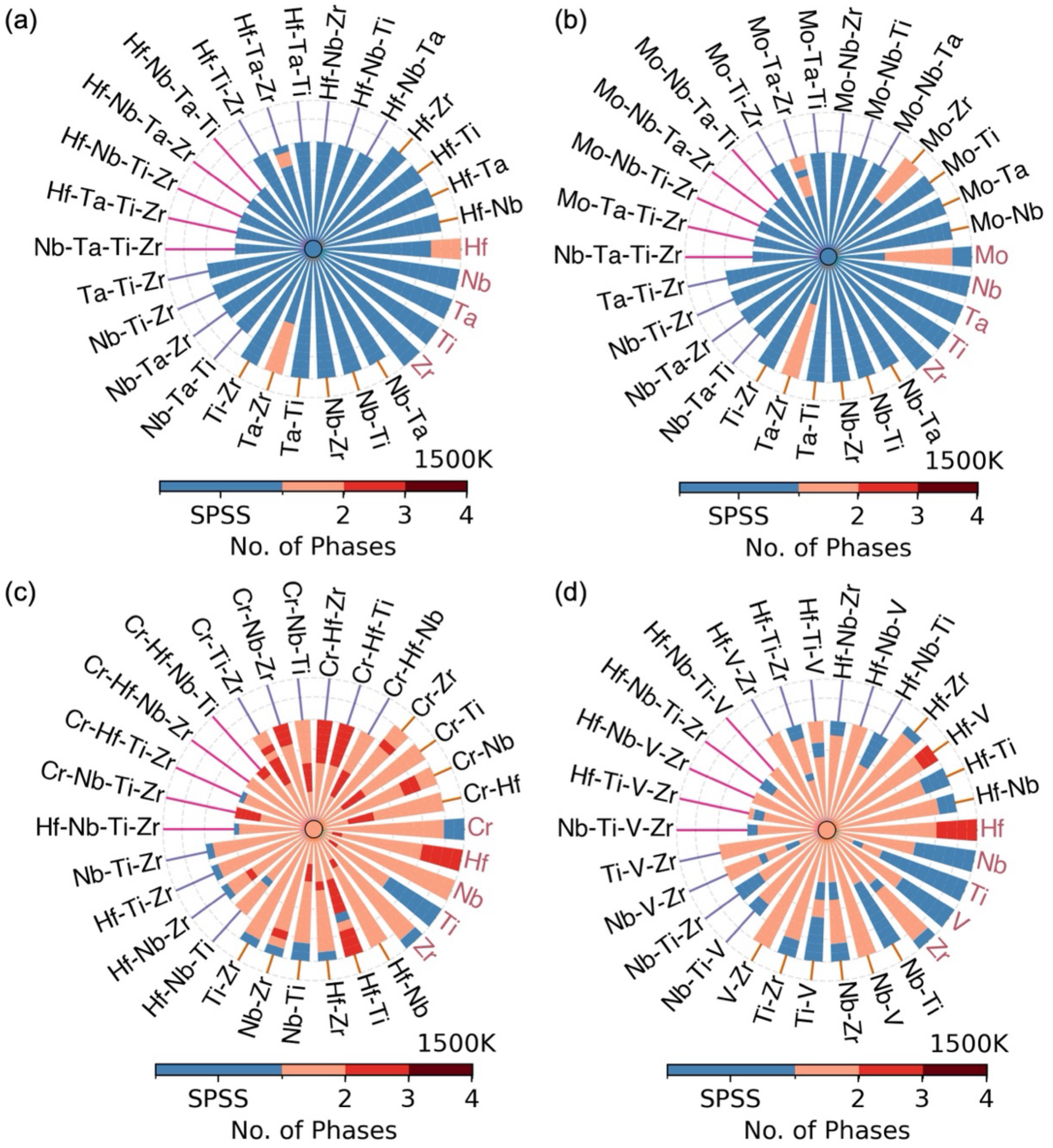

**Fig. 2. SymPlex maps of representative experimentally reported quinary refractory alloy systems.** (a) Hf-Nb-Ta-Ti-Zr, (b) Mo-Nb-Ta-Ti-Zr, (c) Cr-Hf-Nb-Ti-Zr, and (d) Hf-Nb-Ti-V-Zr. The central node corresponds to the equimolar quinary composition, while radial paths connect the equimolar point to unary, binary, and ternary equimolar endpoints. Blue regions indicate stable single-phase solid solution, while red regions indicate the number of phases in equilibrium.

***2.1.2. Quinary systems.*** Fig. 2 extends the same analysis to representative experimentally reported quinary RCCA systems. In comparison to the quaternary maps shown in Fig. 1, quinary phase spaces contain a greater number of lower-order composition paths due to a larger number of combinations. Hf–Nb–Ta–Ti–Zr, as shown in Fig. 2(a), retains a largely connected SPSS region at 1500 K. However, the stability is not completely uniform across the composition space: reduced solid-solution phase fractions appear along selected Hf- and Zr-containing paths. Schuh et al. [17] reported a BCC+HCP phase segregation at 1073 K for the equimolar HfNbTaTiZr alloy, and a single BCC solution above 1173 K, while Senkov et al. [18] reported a BCC solid solution with a cold-worked alloy annealed at 1073 K and 1273 K. In total, out of 10 reported compositions, we correctly predict 9 compositions to show a BCC SPSS across the phase space, well supported by the SymPlex maps.

A similar behavior is observed for the Mo–Nb–Ta–Ti–Zr system shown in Fig. 2(b). The Mo–Nb–Ta–Ti-rich regions remain strongly solid-solution stabilized, while Zr-containing paths show reduction in solid-solution phase fraction. This topology shows that large regions of the surrounding phase space of the equimolar composition remain SPSS-stable at elevated temperatures. Thus, combining our model with the SymPlex map expands the context of the experimental observations from a discrete equimolar composition to a broad composition space.

The Cr-containing Cr–Hf–Nb–Ti–Zr system, Fig. 2(c), shows a much stronger reduction in solid-solution stability compared to the other quinary systems discussed above. The effect of Cr can be visible by the three-phase segregations seen in compositions with Cr > 25 at.%. Fazakas et al. [19] synthesized a CrHfNbTiZr sample and found a three-phase segregation of $NbCr_2$, $HfCr_2$ and a BCC phase at multiple temperatures. At 1100 K, we predict a phase segregation of 13% $TiCr_2$ Laves phase instead of $NbCr_2$ + 17% $HfCr_2$ Laves phase + 70% BCC phase. Whether two intermetallic phases compete to undergo secondary precipitation or prefer to mix in a particular sub-lattice requires further analysis.

Finally, Hf–Nb–Ti–V–Zr, Fig. 2(d), represents a compositionally sensitive quinary system in which the stability varies strongly across the V- and Zr-containing regions. Although several Nb–Ti- and Hf–Nb–Ti-rich paths retain high SPSS fractions, the V–Zr- and Hf/V/Zr-containing regions show reduced stability. In our database we have 6 samples from this system at different compositions. Two of them were heat processed at 873 K and 1073 K [19]. At 873 K, the sample was reported to have a stable BCC SPSS, but at the higher temperature a quinary laves phase (Hf, Zr, Ti, Nb)$V_2$ precipitated out [19]. The model predicts a Laves intermetallic to precipitate out at all temperatures under 1000 K, hence it is possible that kinetics might be playing a role in trapping the metastable SPSS at lower temperatures in the experimental sample. On the contrary, we predict a phase segregation of two BCC solid solutions, 70% Ti-rich phase and 30% V-rich phase at 1073 K. The miscibility temperature of the remaining compositions was predicted to be below their melting temperature, and thus we correctly predict the as-cast samples to be a SPSS. Thus, the experimental systems show a range of different microstructures. Correspondingly, the SymPlex maps exhibit that the solid solution regions in this system are not contiguous and are present as pockets in between phase segregations and intermetallic precipitation.

Taken together, the representative quaternary and quinary SymPlex maps show that the experimentally reported behavior of RCCAs is governed not only by the phase stability at the equimolar composition, but by the size and connectivity of the surrounding miscible region. Robust systems, such as Mo-Nb-Ta-W, retain high solid-solution phase fractions across most composition paths, whereas Cr-, Zr-, and V-containing systems often exhibit localized regions of instability associated with BCC+BCC segregation and ordered precipitates, that can be concentrated along specific lower-order composition pathways. The chemical origins of this behavior, namely strongly positive mixing enthalpies with the other refractories and atomic size-mismatch-driven formation of Laves phases, are discussed in Sections 2.4.2 and 2.4.3. Thus, the SymPlex maps link experimental microstructures to the broader topology of predicted phase fields. To enable systematic comparison across all alloy systems, we next quantify the extent of solid-solution stability.

## 2.2. Predicting stability of SPSS phase fields in RCCAs

The performance of an alloy depends not only on the target composition and microstructure, but also on whether adjacent compositions resist phase segregation during processing or high-temperature service; local deviations in composition can trigger phase segregation, changing the resulting microstructure and properties. We therefore investigate the stability of SPSS phase-fields using a metric that we term as the Percentage Miscible Region (PMR). As shown in the schematic for a ternary system (Fig. 3a), the blue-shaded region represents the ternary space, where all three components are present, while the orange edges represent the constituent binary spaces. We define an alloy *system* as the entire composition range encompassing all constituent phases (including unaries and binaries), whereas a *space* is strictly limited to compositions where all components are present simultaneously (having non-zero mole fractions). This hierarchical definition is extended to quaternary and quinary systems. To calculate the PMR, we sampled over 100 individual compositions uniformly distributed within each *space* (so to calculate the PMR for a ternary space, we exclude the binary and unary alloy compositions), for every alloy system, to determine the percentage of compositions exhibiting a stable SPSS.

We use a strict threshold to define PMR: at a given temperature, any composition with an SPSS phase fraction less than unity is classified as phase segregated. We omit soft cutoffs because even minor phase fractions (<1%), like nanoprecipitates, can impact mechanical or functional properties [20]. However, as noted in the previous section, thermodynamically stable phases may not form experimentally in some cases due to kinetic limitations tied to processing conditions.

We define three ranges of PMR: low (<25%), mid (25–65%), and high (>65%). High-PMR systems are predominantly single-phase solid solutions (SPSS), low-PMR systems are dominated by phase segregation, and mid-PMR systems contain coexisting SPSS and multi-phase domains. Mid-PMR alloys are particularly advantageous for design as they enable controlled precipitation strengthening while preventing the high intermetallic volume fractions that impair ductility. We discuss the implications of mid-range and high-range PMR in the following.

***2.2.1. Mid-range PMR systems.*** The utility of mid-range PMR systems is illustrated using the Hf-Nb-Ta-W system, shown in Fig. 3(b). The Hf-Nb-Ta-W system is not in our experimental database and is yet to be studied for its properties. The SymPlex map [12] for this quaternary system at 1500 K displays compositions containing >25 atomic % of W (i.e., exceeding the W content of equimolar HfNbTaW alloy) in the upper half and compositions with <25 atomic % of W in the lower half. At 1500 K, the Hf-Nb-Ta-W system exhibits a PMR of 55%, indicating that approximately half of the sampled composition space stabilizes as an SPSS. This bifurcation in stability is primarily driven by W; the formation of a stable $HfW_2$ Laves phase [21] destabilizes the solid solution. Accordingly, W-rich compositions, as seen from the paths in the upper half of Fig. 3(b), do not stabilize in an SPSS. Thus, in Hf-Nb-Ta-W, varying the relative Hf-W and Nb-Ta content can trigger the precipitation of Laves phases.

Similarly, other mid-range PMR systems offer microstructural sensitivity to compositional changes, enabling tuning of mechanical properties. Because solid solution matrices reinforced with ordered precipitates can exhibit enhanced mechanical performance [22], mid-range PMR systems provide a strategic pathway to optimize alloy functionality. Also, knowing that systems fall in a mid-range PMR allows for developing strategies to safeguard them against compositional fluctuations, diffusion and radiation damage, which could trigger large changes in the phase compositions at high operating temperatures. For example, the synthesized alloy composition can be chosen such that they fall deep within an SPSS region, rather than close to the boundary or appropriate heat treatments can be applied so as to ensure that compositional fluctuations do not lead to deleterious phase segregations. Hence, PMR can be a very useful metric for designing alloys which can be applied on an industrial scale. These metrics are method-invariant, and thus can be adopted for use with other phase diagram prediction methods too.

***2.2.2. High-range PMR systems.*** We also briefly discuss systems having high-range PMR. The Mo-Ti-V-W system, shown in Fig. 3(c) is a high-PMR system, with a PMR of 100%, and is yet to be studied for its properties to the best of our knowledge. Such systems offer distinct advantages, particularly for high-temperature and extreme-environment applications. In such environments, local compositional

perturbations—driven by diffusion, chemical segregation, or oxidation—are common [9, 23, 24]. As stated earlier, an alloy system with a high-range PMR possesses SPSS that are inherently robust against these fluctuations, as it is more likely to maintain a stable solid solution across a broad stoichiometric range. This compositional tolerance can be critical for preserving mechanical strength, thermal stability, and resistance to deleterious phase segregation under extreme operating conditions such as turbine blades in jet engines. We note that although oxide and defect formation also impact alloy performance, assessing these factors is beyond the scope of this study.

Given that only 50 of the 252 possible quaternary and quinary RHEA systems have been experimentally reported, to the best of our knowledge, as tabulated in Table 1, a vast combinatorial space remains unexplored. Fig. 3(d) illustrates the predicted distribution of PMR at 1500 K across the 84 ternary, 126 quaternary, and 126 quinary refractory alloy systems. We identify 14 ternary, 21 quaternary, and 21 quinary systems within the mid-range PMR regime. High-range PMR is observed in 29 ternary, 29 quaternary systems, and 16 quinary systems. These largely unexplored systems offer significant opportunities for discovering alloys with tailored functionalities. A list of mid and high-range PMR systems is provided in Appendix B and an extended list is available in the associated public repository [15].

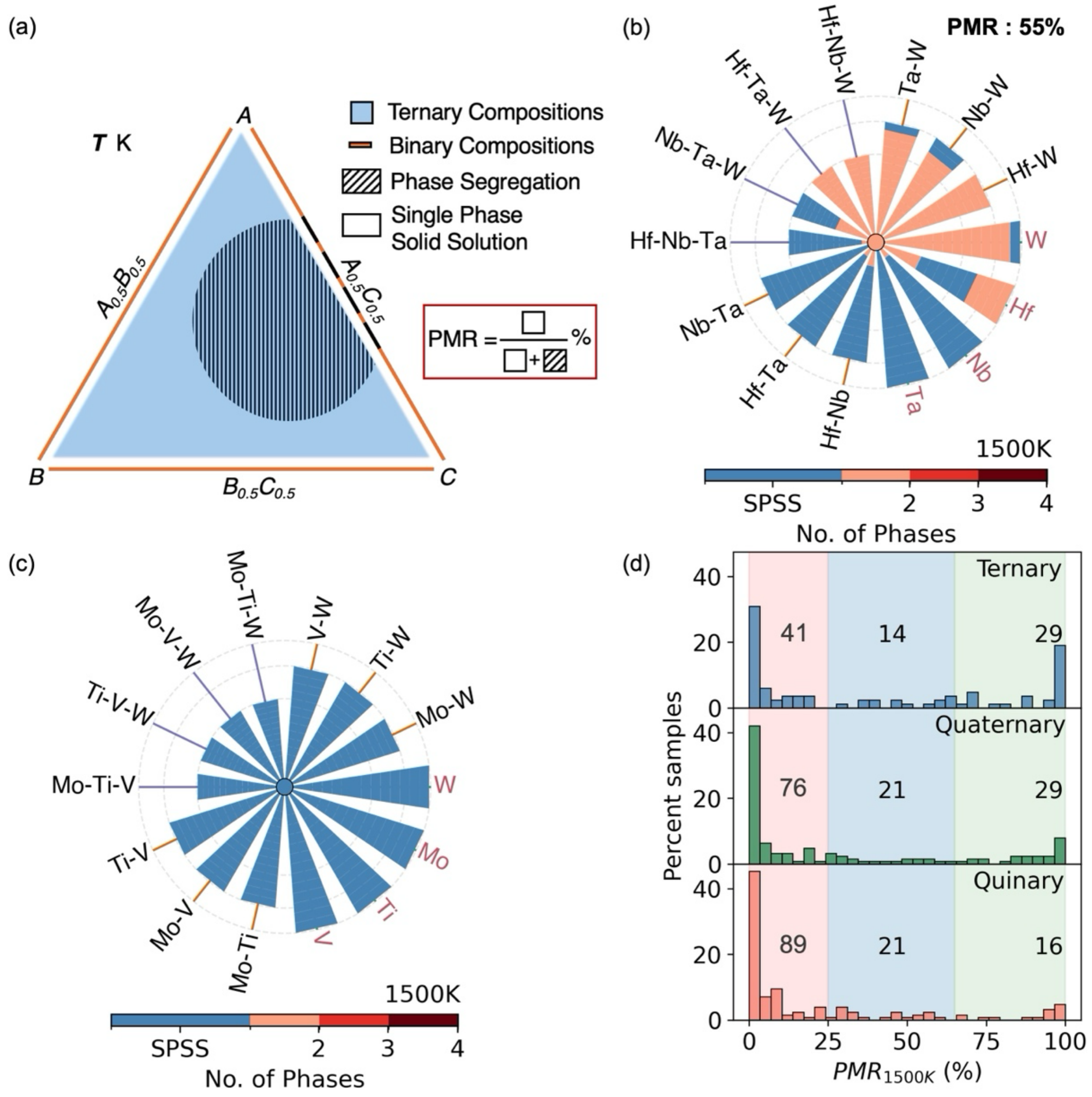

**Fig. 3. Percentage Miscible Region (PMR) as a metric for assessing the compositional stability of RCCAs.** (a) Schematic illustration of PMR calculation in a ternary system. Blue shading denotes the ternary phase space, orange edges represent binary subsystems, and vertices indicate unary elements; shaded regions within the triangle represent phase-segregated domains. (b) SymPlex map of the Hf-Nb-Ta-W system at 1500 K, exemplifying a mid-range PMR system (55%). (c) SymPlex maps of the Mo-Ti-V-W system at 1500 K, exemplifying a high-range PMR system (100%). (d) Distribution of predicted PMRs at 1500 K for ternary (top), quaternary (middle), and quinary (bottom) systems. Low-, mid-, and high-range PMR regimes are highlighted in red, blue, and green, respectively.

## 2.3. Topology of miscibility gaps in higher-order RCCAs

To understand the origins of solid solution stability in complex systems, we analyze the topology of miscible regions by using quinary alloys as test systems. We decompose the multinary alloy phase spaces into their constituent lower-order subsystems. This hierarchical division is illustrated for a quaternary system in Fig. 4(a), where the phase space is partitioned into three distinct domains: 1) the quaternary interior (high configurational entropy), 2) the four constituent ternary systems forming the faces of the Gibbs tetrahedron, and 3) the six constituent binary systems forming its edges (lowest configurational entropy). Because these three geometric domains: the interior, faces, and edges, do not overlap in composition space, they are mutually exclusive. By extending this logic to quinary systems, we can evaluate how miscibility *propagates* from lower-entropy binary and ternary boundaries into the high-entropy quinary interior.

***2.3.1. Propagation of miscibility across phase space.*** We depict the *propagation* of the miscible region across the high-dimensional phase space using the Sankey flow diagram shown in Fig. 4(b), which correlates the PMR of each quinary interior with the average PMR of its constituent lower-order subsystems. The PMR distributions are discretized into three bins indicating the three PMR ranges outlined in the previous section, with the flow lines connecting the high-range PMR regimes highlighted in green, between mid-range PMRs highlighted in blue, and the rest in red. The flow lines indicated by red arrows reveal a direct correlation: *quinary systems exhibiting high compositional stability are almost exclusively derived from subsystems that also possess high average PMRs.* While this relationship may appear intuitive, it uncovers a constraint in alloy design.

***2.3.2. Obtaining high-range PMR in higher-order alloys.*** Because the relationship is defined in terms of an average, achieving a high quinary PMR requires that all constituent subsystems, particularly the binaries, are largely miscible. Even a single immiscible binary with a low PMR can substantially reduce the overall average, ultimately destabilizing and collapsing the single-phase region in the quinary space. The flow diagram in Fig. 4(b) depicts this sensitivity in the stability of higher-order solid solutions. This finding carries implications for the role of configurational entropy in stabilizing solid solutions; it suggests that the increase in entropy gained by adding more elements is often insufficient to stabilize a solid solution if the underlying binary interactions are strongly segregating. To assess how much unfavorable binary behavior can be accommodated in a compositionally stable quinary system, we calculated the number of high-PMR binary subspaces associated with each high-PMR quinary system. Each quinary alloy contains 10 constituent binary subspaces. Nearly all high-PMR quinary systems—15 out of 16—contain ≥6 high-PMR binary subspaces. Moreover, the probability of obtaining a high quinary PMR increases to 65% when at least 7 out of the 10 constituent binary subspaces have high PMRs. Thus, a quinary system can retain a broad SPSS composition space when 2 or 3 of its constituent binary subspaces do not belong to the high-PMR category. Nevertheless, the results indicate that majority of the constituent binary subspaces must generally exhibit high PMRs to obtain a quinary solid solution with robust compositional stability.

***2.3.3. Obtaining mid and low-range PMRs in higher-order alloys.*** The Sankey diagram in Fig. 4(b) further reveals a transition within the mid-range quinary PMR regime. Most quinary systems in the mid-range PMR category originate from binary subsystems with a high average PMR, indicated in Fig. 4(b) by the black arrows. Additionally, we plot the mid-range quinary PMR systems associating them with their binary subspaces in Fig. 8 from Appendix C. Quantitatively, out of the 21 systems classified to be mid-range PMR in the quinary space, 14 (66.7%) systems are associated with high average binary PMR.

The analysis above underscores the central role of binary interactions in governing the miscibility of higher order alloys and highlights the effectiveness of binary mixing enthalpies as descriptors for multinary alloys. It further shows that even localized miscibility in higher-order space, represented by mid-PMR alloys, requires constituent binaries to be largely miscible. Most quinary RCCAs have PMRs ≤ 25%, consistent with previous observations [25, 26] that higher-order systems show reduced miscibility compared to lower-order subsystems. *Collectively, these results establish a clear design rule: quinary RCCAs with large miscible regions can be created by carefully choosing constituent binary alloys, each possessing high miscibility*.

***2.3.4. Propagation of miscibility gaps in higher-order alloys.*** The propagation of miscibility described above also hints at the physical topology of the miscibility gap surrounding equimolar quinary compositions. This is visualized in Fig. 4(c) via a schematic of a 2D projection of the high-dimensional phase space, where the center represents the equimolar composition and the configurational entropy generally increases from the edges toward the center. At a certain temperature (indicated by the dashed black line), if the quinary interior were miscible, while its lower-order boundaries remained immiscible, the phase diagram would resemble the topology shown in the top panel of Fig. 4(c). However, the Sankey diagram in Fig. 4(b) demonstrates that a stable solid solution in the interior quinary space is almost always accompanied by stability in the lower-order regions. Isolated stabilization of the quinary interior is therefore uncommon. Consequently, the more representative topology is that shown in the bottom panel of Fig. 4(c), where the miscibility region extends from the high-entropy interior into the lower-entropy subsystems. It is to be noted that the miscibility gap schematics collapse a five-dimensional space into 2D, and thus the phase boundaries are not necessarily as smooth as indicated. The schematic is only meant to depict that the quinary compositions stabilize as a SPSS at higher temperatures than their lower order subsystems. We quantified the topology across all 126 equimolar quinary systems, as shown in Fig. 9 in Appendix C. The median increase in $T_{misc}$ of the quinary *relative* to the average of its binary, ternary, and quaternary subsystems is +710 K, +335 K, and +110 K, respectively. The positive values indicate that the $T_{misc}$ of the subsystems lie at lower temperatures than the quinary peak, and the gap narrows as the subsystem order approaches five, aligning with the miscible boundary shown in the schematic in Fig. 4(c).

***2.3.5. Equimolar stability as an indicator of regional miscibility and high PMR.*** A key topological feature of the miscibility gap, shown in the bottom panel of Fig. 4(c), is that the equimolar composition exhibits the highest miscibility temperature. Consequently, at any temperature where the equimolar point is miscible, a substantial fraction of the surrounding region is also likely to be miscible (indicated by the yellow dashed line). *Equimolar SPSS stability can therefore serve as an indicator of broad regions of miscibility and therefore high PMR.*

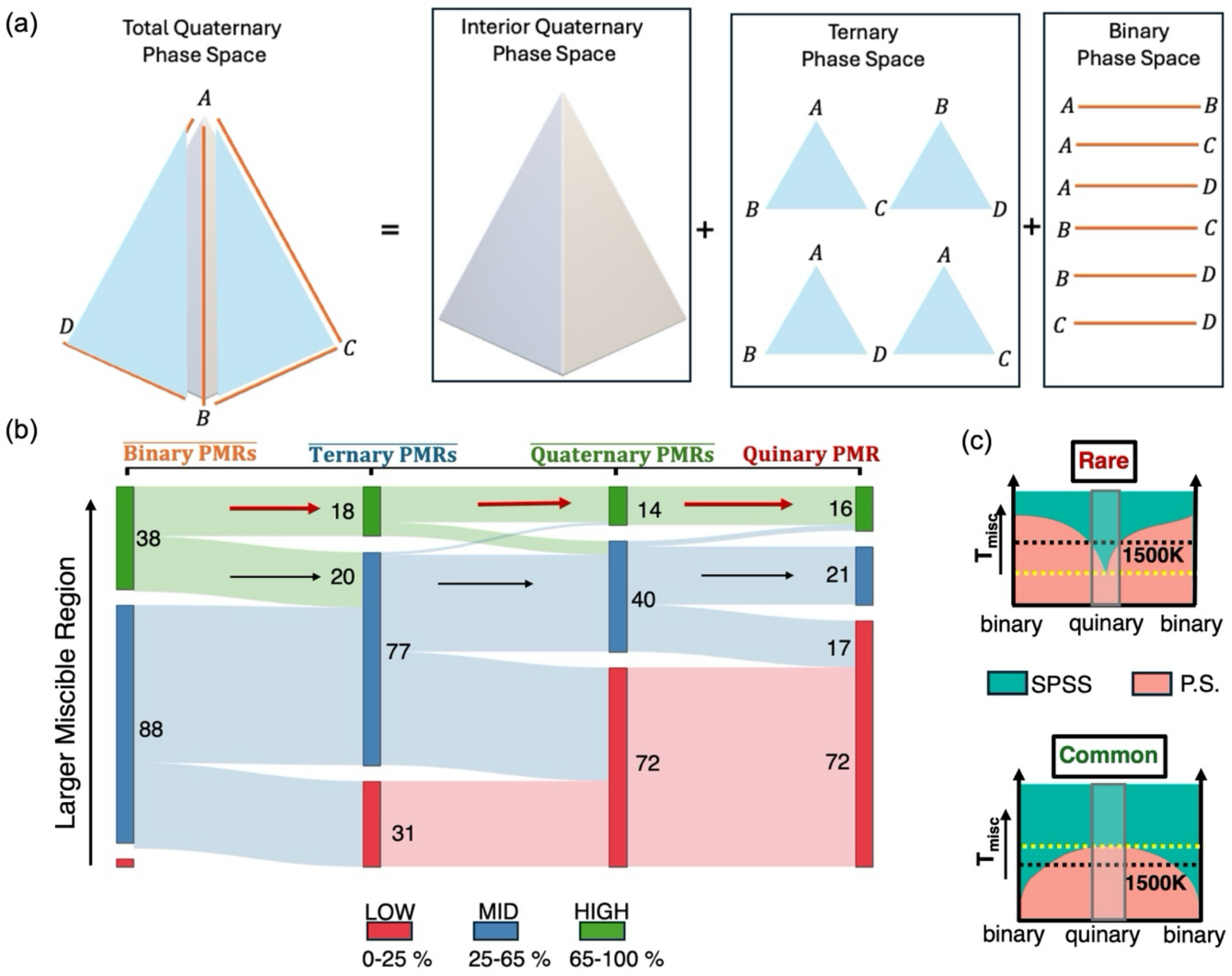

**Fig. 4. Topology of miscible regions in higher-order RCCAs.** (a) Schematic of a Gibbs tetrahedron for a quaternary alloy system (ABCD), partitioned into quaternary compositions (interior), four constituent ternary subsystems (faces), and six constituent binary subsystems (edges). (b) Sankey diagram correlating the PMR of the quinary space with the average PMRs of its constituent lower-order subsystems. The PMR distributions are categorized into low, mid and high ranges. (c) Schematic representing the topology of the quinary miscibility gap within a 2D-projected compositional space. Configurational entropy increases towards the center in both the cases. The horizontal, dotted black line represents an isotherm at 1500 K, at which the analysis was performed. The horizontal, dotted yellow line represents the temperature at which the SPSS is stable for the equimolar composition.

## 2.4. Effect of chemistry on the solid solution stability in RCCAs

Building on the previous section, we examine the role of chemistry, specifically, whether particular elements promote or hinder SPSS stability in RCCAs. We probe the relationship between the enthalpy of mixing, $\Delta H_{mix}$, and the normalized miscibility temperature, $\frac{T_{misc}}{T_{melt}} = \chi_T$, as shown in Fig. 5(a) for equimolar ternary, quaternary, and quinary alloys. The analysis is restricted to equimolar compositions to isolate the effect of elements, rather than their composition. A horizontal dashed line at $\chi_T = 1$ denotes the limit where an alloy becomes miscible at the average of the melting temperature of the constituent elements. while a threshold of $\chi_T = 0.9$ is adopted to define solid solution stability. The average is only a proxy for the alloy melting temperature, so the lowest melting temperature among the constituent elements was also tested as a conservative bound. This lowers $T_{melt}$ by 20% on average and reduces the number of systems with $\chi_T \leq 0.9$ from 209 to 143 of the 336 ternary, quaternary, and quinary systems, but the near-linear correlation between $\Delta H_{mix}$ and $\chi_T$ is unchanged, as is its dependence on the sign of $\Delta H_{mix}$.

We observe a near-linear correlation between $\Delta H_{mix}$ and $\chi_T$. In many metallic systems, as reported by Chen et al. [11], the formation energy of intermetallic phases is often directly proportional to $\Delta H_{mix}$, meaning that as $\Delta H_{mix}$ of the BCC solid solution becomes more negative, ordered intermetallics are favored over a solid solution. Positive values of $\Delta H_{mix}$ favor phase segregation and lead to solid solution instability. Therefore, for both highly negative $\Delta H_{mix}$ and positive $\Delta H_{mix}$ values, the solid solution should be unstable, i.e., $\chi_T \geq 1$. For the refractory alloys studied here, we find that highly positive $\Delta H_{mix}$ values correlate with $\chi_T \geq 1$, but, for $\Delta H_{mix} < 0$, the alloys become miscible, in contrast to the findings by Chen et al. [11, 27]. This is because the competing intermetallics for refractory systems are predominantly Laves phases. As established in Part I of this work, stable binary Laves phases correlate with positive binary $\Delta H_{mix}$. *Thus, in RCCAs, highly positive $\Delta H_{mix}$ values often indicate a strong propensity for the precipitation of Laves phases, and high propensity for phase segregation.* This contrast arises because Laves phases are atomic size-driven rather than bonding-driven compounds, as discussed in Part I. A large and a small atom pack efficiently in the Laves structure, which gives the compound a negative formation energy, whereas the same size mismatch strains the BCC lattice and produces a positive $\Delta H_{mix}$ for the solid solution. The sign of the BCC $\Delta H_{mix}$ and the stability of the competing intermetallic are therefore two separate quantities, and for Laves formers they have opposite signs. This is why the intermetallic-forming systems in Fig. 5(a) appear at positive $\Delta H_{mix}$, opposite to where the conventional rule would place them. The conventional rule is not triggered on the negative side for these elements, leading to a competing, isostructural ordered compound. Thus, the prediction of stable SPSS in RCCAs is most accurately achieved through the explicit construction of phase diagrams.

***2.4.1. Ti, Mo, Nb, and Ta as solid solution stabilizers.*** To isolate the influence of individual elements, we assess the sensitivity of these systems by calculating the shift in $T_{misc}$ when a fifth element is added to a quaternary equimolar base, as shown schematically in Fig. 5(b). Specifically, for each of the nine refractory elements, we quantified the change across all possible quaternary-to-quinary equimolar transitions. The impact of an alloying addition is intrinsically tied to its chemical environment; the same element can either destabilize or stabilize an SPSS alloy depending on the composition of the base alloy. This can be seen in Fig. 5(c), with each element contributing to both increase and decrease in $T_{misc}$ depending on the quaternary base composition. Nevertheless, clear trends emerge: Ti, Mo, Nb, and Ta consistently act as SPSS stabilizers, generally lowering $T_{misc}$ regardless of the base quaternary system. This behavior stems from their relatively moderate and similar interactions with other refractory metals in the BCC phase. These moderate interactions arise from the atomic radii of Ti, Mo, Nb, and Ta; all lie within about 4 % of the mean radius of the nine refractory elements, and thus these elements form an unstable or weakly stable Laves phase. Chen et al. [11] also found Mo and Nb to be present in many stable SPSS of quinary alloys. However, they didn't account for the HCP to BCC phase transition in Ti, which leads to the stabilization of many Ti-containing BCC SPSS, as observed here.

***2.4.2. Zr and Cr as solid solutions destabilizers.*** In sharp contrast, Zr and Cr act as potent destabilizers due to their strong tendency to form Laves-phase. In section 2.2, the SymPlex maps of experimentally reported systems highlights the role of Cr and Zr as destabilizers of solid solutions. Cr, in particular, forms stable Laves phases such as $ZrCr_2$, $HfCr_2$, $TaCr_2$, $TiCr_2$ and $NbCr_2$. Zr and Cr sit at opposite ends of the size range, about ±11 % around the mean radius of the nine elements, so they pack efficiently into the Laves structure with the other elements. Furthermore, Cr-W and Cr-Mo binaries show strong miscibility gaps on account of their positive $\Delta H_{mix}$.

***2.4.3. Ambivalent nature of Hf, V and W.*** Perhaps more significant for alloy design is the "ambivalent" nature of Hf, V, and W. These elements can either stabilize or destabilize the solid solution depending on the specific chemistry they are introduced into. For example, V is sensitive to the presence of Ta, as their mutual interaction can trigger the formation of a stable $TaV_2$ Laves phase. Similarly, Hf forms solid solutions when paired with Ti, Zr, Nb and Ta, but precipitates into a Laves phase when paired with Mo or W, as discussed in Part I. W too forms a SPSS when paired with V, Mo, Nb and Ta; however, it can precipitate out Laves phase when paired with Hf and Zr. These results demonstrate that the impact of Hf, W, and V on RCCA microstructures cannot be evaluated in isolation; rather, their roles must be understood within the context of the specific chemical environment. Importantly, they can be paired with other carefully chosen elements to create mid-range PMR systems. Indeed, the previously discussed Hf-Nb-Ta-W alloy system in Fig. 3(b) serves as an illustrative example of pairing Hf with Nb/Ta and also with W to result in a PMR of 55%. Thus, these elements can be used flexibly depending on the functionality and can be valuable for tailoring the microstructure while improving other desired properties like elastic moduli or melting temperature.

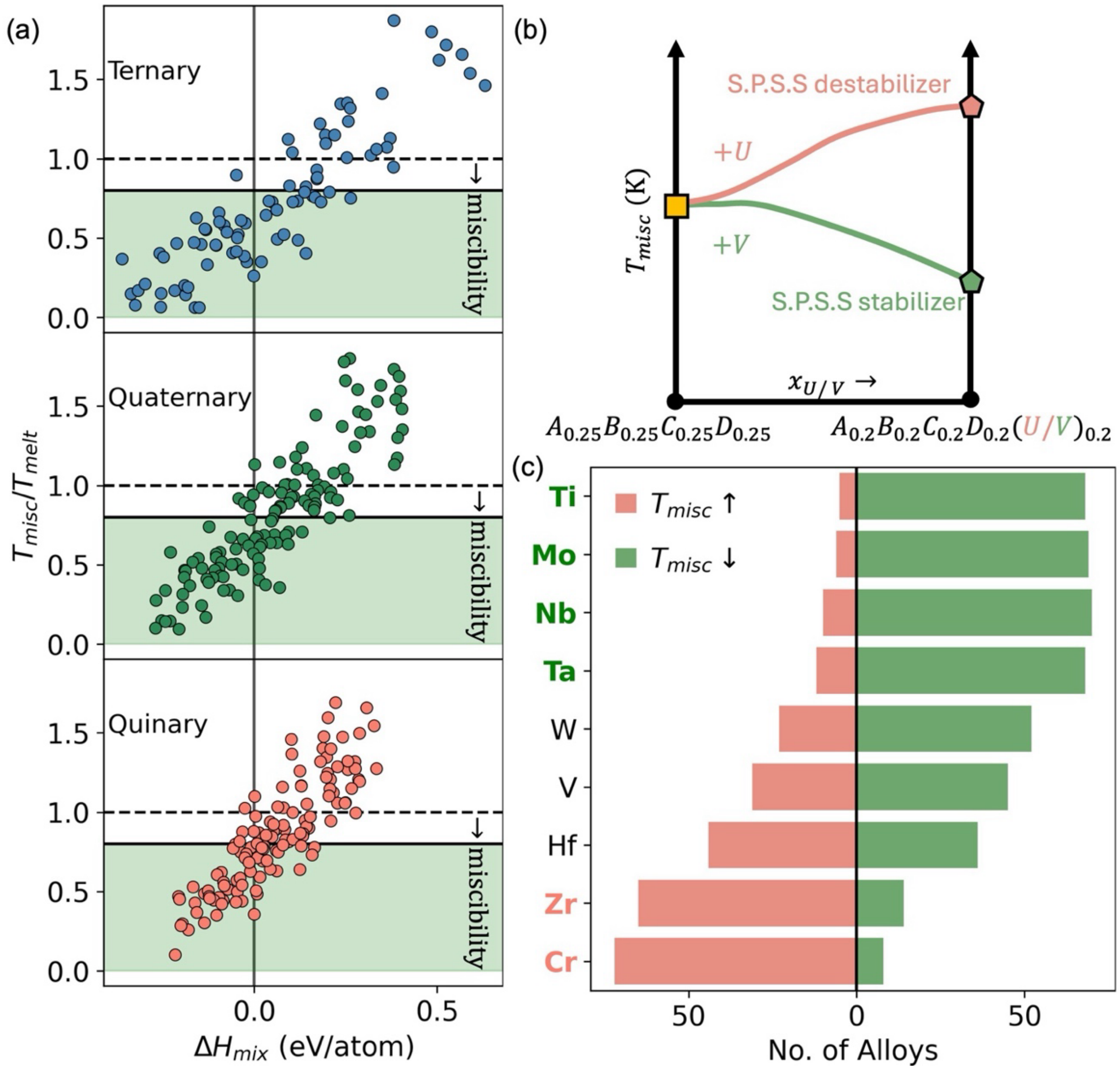

**Fig. 5. Influence of chemical composition on the miscibility temperature ($T_{misc}$) of equimolar RCCAs.** (a) Enthalpy of mixing ($\Delta H_{mix}$) plotted against $T_{misc}$ normalized by the average melting temperature ($T_{melt}$). The horizontal dashed line at $\frac{T_{misc}}{T_{melt}} = 1$ denotes the threshold where the composition is considered immiscible; a ratio of 0.9 is utilized as the practical cutoff for solid solution stability. (b) Schematic illustrating shifts in $T_{misc}$ upon the addition of a fifth element to an equimolar quaternary alloy, highlighting cases of increase (red) and decrease (green) in $T_{misc}$. (c) Statistical distribution showing the number of instances where the addition of a fifth element to quaternary equimolar alloys results in an increase (red) or decrease (green) in $T_{misc}$.

## 2.5. Effect of temperature on the stabilization of solid solution in RCCAs

Finally, we examine the influence of temperature on the stabilization of SPSS in RCCAs by tracking the number of phases with temperature, in expanding the solid solution regions in RCCAs.

***2.5.1. Number of equilibrium phases at different temperatures.*** We first analyze the equilibrium phase count (*f*) at different temperatures collated for 100 compositions in each alloy system, starting with 500 K in Fig. 6(a). At 500 K, 42% and 46% of the ternaries exhibit two- or three-phase regions, respectively. Thus, only ~10% of the ternaries are SPSS at 500 K. 75.2% of the quaternary compositions show three- or four-phase equilibrium, while 62% of the quinaries reach four- or five-phase segregation. Thus, the general complexity of the microstructures continues to increase with the increase in order. At 500 K, however, substitutional diffusion in refractory BCC alloys is negligible for any practical timescale, so these four- and five-phase equilibria will not form in practice. The 500 K maps simply show the thermodynamic driving force. Despite this initial complexity, the higher entropic drive in quinary systems at 1500 K effectively collapses these multinary segregations into predominantly two-phase microstructures.

At 1500 K, 40% of the ternary compositions show stable SPSS, while 40% exhibit two-phase segregation. In quaternary and quinary systems, two-phase segregation predominates (~45% each), whereas multiphase microstructures (three or more phases) account for 26% of quaternaries and 30% of quinaries. These systems are thus useful to obtain complex microstructures consisting of a combination of multiple ordered and disordered phases, and can show enhanced properties, as has been shown recently by Zhang et al. [4].

The prevalence of two-phase microstructures aligns with findings by Luan et al. [28], who attributed it to the unidirectional nature of spinodal decomposition in higher-order MPEAs. They analyzed the curvature of solid solution mixing enthalpies in MPEAs and found that a single direction of elemental partitioning dominated over the other directions. For example, a quaternary *ABCD* might split into *AB* + *CD*, rather than segregate into *A*+*B*+*C*+*D*. While their work focused only on a single solid solution for each system, our work incorporates intermetallics and multiple phases into the free-energy calculations. Notably, even with the inclusion of intermetallics, our results demonstrate that two-phase segregation remains the most prominent form of phase instability in higher order RCCAs at elevated temperatures. Thus, RCCAs, even with increased chemical complexity, are expected to show two-phase microstructures on average. These two-phase microstructures have been extensively studied in binary and ternary systems and the knowledge gained from these studies can still be used for MPEAs, even with increased chemical complexity.

***2.5.2. Expansion of solid solution regions with increasing temperature.*** Next, we analyze how the solid solution regions in RCCAs expand/contract with change in temperature. The increase in temperature, leads to the entropy part of the Gibbs' free energy to stabilize solid solutions. Note that within the quinary phase space the configurational entropy is fixed with composition and it is the increase in the temperature which leads to a competition with the enthalpy terms. Hence an expansion of the SPSS compositional region will occur when heating an alloy from 500 K to 1500 K, as shown using two SymPlex maps of Hf-Ta-Ti-W in Fig. 6(c). The dark blue regions indicate SPSS. Note that at 500 K, kinetic trapping and limitations in reaching equilibrium are much pronounced, and thus any comparisons to experimental samples at this temperature depends on annealing time and quenching rates. At 500 K, Hf-Ta-Ti-W consists of minimal SPSS regions, but at 1500 K, more than half the entire region exhibits a SPSS. The PMR thus rises from 0% at 500 K to 65% at 1500 K. We quantify the expansion of the miscible region for all the alloy systems by calculating the PMR at 500 K and 1500 K, shown as violin plots in the left and middle panels of Fig. 6(d). Black lines, connect the medians of the two distributions. We next quantify the temperature-driven expansion of the SPSS region using the system-wise difference of PMRs at 500 K and 1500 K, ΔPMR, shown as a violin plot in the rightmost panel of Fig. 6(d). At 500 K, the PMR distributions are strongly concentrated near zero for ternary, quaternary, and quinary systems, indicating that large portions of the refractory composition space are phase segregated at low temperatures. Upon heating to 1500 K, the miscible region expands for all alloy orders, but the magnitude of this expansion is order dependent. The average PMR at 1500 K falls from 41% for the ternary systems to 29% for quaternaries to 21% for quinary systems. Thus, ternaries show the strongest expansion, with the miscible regions on average *increasing* by 30% compared to 25% and 20% for quaternaries and quinaries, respectively. It should also be noted that the ΔPMR for quaternaries and quinaries consist of a short whisker extending below 0, indicating a minority of systems — 4 for quaternaries and 3 for quinaries — show a contraction in miscible region. The list of PMRs at different temperatures for binary to quinary alloys studied in this work is presented in the public repository [15].

Collectively, these results show that while configurational entropy promotes phase-field simplification and expands the SPSS region at elevated temperatures, its ability to overcome adverse chemical interactions becomes progressively limited as the number of components increases. Consequently, reinforcing our earlier findings, *higher-order RCCAs cannot be stabilized by entropy alone; rather, their stability remains tightly constrained by the chemistry of their lower-order subsystems.*

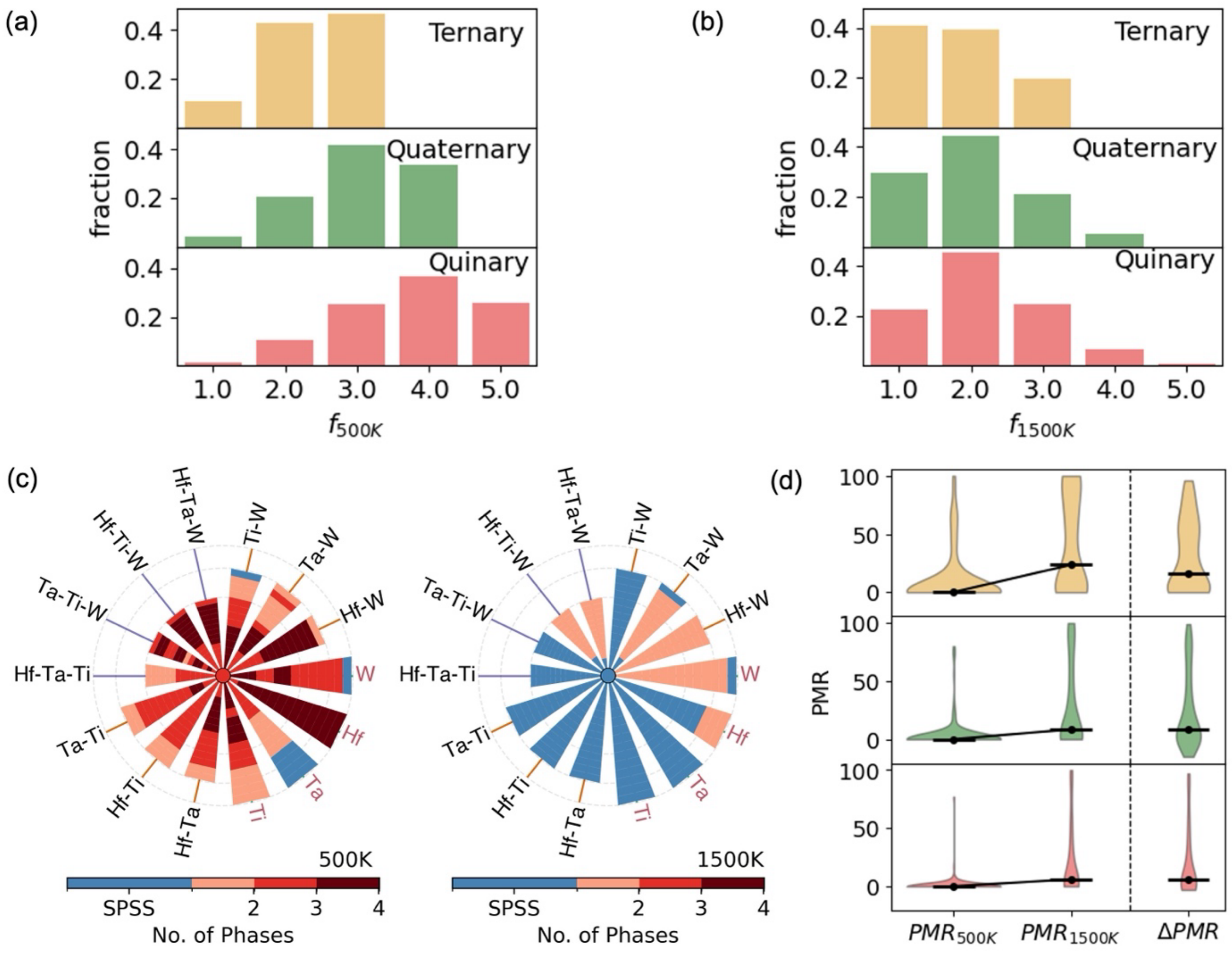

**Fig. 6. Influence of temperature on the stabilization of solid solutions in RCCAs.** Frequency distribution of the number of phases (*f*) present at (a) 500 K and (b) 1500 K for ternary, quaternary, and quinary systems. (c) SymPlex maps of the quaternary Hf-Ta-Ti-W system at 500 K and 1500 K, showing the expansion of the miscible region with increasing temperature. (d) Violin plots quantifying the expansion of the SPSS regions using PMR values at 500 K, 1500 K, and the difference in each system's PMR between 500 and 1500 K. Black lines indicate the medians of the three distributions.

## 3. Conclusion

In this two-part study, we have developed and validated a first principles-based predictive framework for the phase stability of RCCAs. While Part I established the model and benchmarked its predictive power for binary and ternary systems, the second part extends the model to refractory multinary systems, demonstrating an 80% accuracy against a diverse experimental dataset of 109 compositions across varied processing routes.

We introduced the *percentage miscible region* as a metric to quantify the compositional stability of solid solutions. Our topological analysis reveals that the stability of higher-order systems is intrinsically tied to their constituent binaries. Quinary alloys with high compositional stability—useful for applications where local fluctuations in composition can occur—require the selection of high-PMR binary subsystems. Additionally, we identified that mid-range PMR systems (25–65%) are underexplored in the literature and offer a space to obtain tunable microstructures. These systems, listed in Appendix B, typically consist of a strategic mix of miscible and intermetallic-forming binary pairs, allowing for the precise control of precipitate-strengthened matrices.

We categorized refractory elements into phase stabilizers (Ti, Mo, Nb, Ta) and destabilizers (Zr, Cr), with Hf, V, and W exhibiting more nuanced behavior depending on the base alloy composition. Our investigation into temperature effects confirms that while increasing configurational entropy reduces the number of equilibrium phases and expands the solid solution regions, two-phase segregation remains the most prominent form of instability in higher-order refractory alloys.

By mapping the unexplored combinatorial space of RCCAs, this work provides a rigorous theoretical foundation for moving beyond equimolar compositions toward the targeted design of high-performance alloys with tailored microstructures and enhanced functional stability.

## Data Availability

A python-based package to analyze the MPEA phase diagrams predicted in this work has been provided in https://github.com/Materials-Modelling-Microscopy/RAPTor along with nearly 400 TDB files. Code to plot phase diagrams, compute miscibility temperatures and parse TDB files are also provided. The DFT calculations are provided in a publicly available Zenodo repository [15]. A web interface is also made available to obtain predicted phase diagrams and related analysis for the alloys studied in this work at https://raptor.engr.wustl.edu

## Declaration of Competing Interest

There are no competing interests to declare.

## Acknowledgements

This work was partially supported by the Schmidt Family Foundation (P.O., N.C., J.C., A.C., R.M.) and the National Science Foundation through award # DMR-2145797 (P.O., R.M.). This work used computational resources through allocation DMR160007 from the Advanced Cyberinfrastructure Coordination Ecosystem: Services & Support (ACCESS) program, which is supported by NSF awards #2138259, #2138286, #2138307, #2137603, and #2138296.

## Appendix A. Comparison of model predictions with experimental reports

**Table 1. Comparison of predicted phase stability against experimental results for 109 quaternary, quinary, and senary refractory complex concentrated alloys (RCCAs).** The table summarizes the accuracy of the model across diverse processing routes, As-cast (Cast), Heat Processed (Anneal) and Additive Manufacturing (AM), distinguishing between single-phase solid solutions and multi-phase segregated microstructures. Entries in bold are misclassifications. Thick, black lines demarcate alloy systems of different number of components.

| Composition | Expt. Reported Phases | Processing Method | $T_{proc}$ | $T_{misc}$ | Ref. |
|---|---|---|---|---|---|
| **CrMoNbTi** | **BCC** | **Anneal** | **1673** | **2600** | [16] |
| **CrMoNbV** | **BCC** | **AM** | **2180** | **3000** | [29] |
| **CrMoTaTi** | **BCC** | **Cast** | **1941** | **2300** | [30] |
| CrMoTaTi | BCC+Laves | Anneal | 1673 | 2300 | [16] |
| CrNbTaTi | BCC+Sec. | Anneal | 1673 | 2300 | [30] |
| **CrNbTaTi** | **BCC** | **Cast** | **1941** | **2300** | [30] |
| CrNbTiW | BCC+Sec. | Anneal | 1673 | 2500 | [30] |
| **CrNbTiW** | **BCC** | **Cast** | **1941** | **2500** | [30] |
| $Cr_3NbTiZr$ | BCC+Laves | Anneal | 1473 | 3200 | [31] |
| CrNbTiZr | BCC+Sec. | Anneal | 1473 | 3500 | [32] |
| CrTaVW | BCC+Laves | Powder | 1773 | 2200 | [33] |
| HfNbTaTi | BCC | Anneal | 1673 | 600 | [16] |
| **HfNbTaZr** | **BCC+HCP** | **Anneal** | **2073** | **1100** | [34] |
| HfNbTaZr | BCC | Cast | 2128 | 1100 | [34] |
| **$Hf_{28}Nb_{28}Ti_{28}Zr_{16}$** | **BCC+Sec.** | **AM** | **1941** | **700** | **[3]** |
| HfNbTiZr | BCC | Anneal | 1073 | 800 | [35] |
| $Hf_{26}Nb_{26}Ti_{26}Zr_{18}$ | BCC | AM | 1941 | 700 | [3] |
| HfNbTiZr | BCC | AM | 1941 | 800 | [3] |
| **$Hf_{29}Ta_{12}Ti_{29}Zr_{29}$** | **BCC+HCP** | **Cast** | **1941** | **1200** | [36] |
| **$Hf_{29}Ta_{14}Ti_{29}Zr_{29}$** | **BCC+HCP** | **Cast** | **1941** | **1200** | [36] |
| **$Hf_{28}Ta_{17}Ti_{28}Zr_{28}$** | **BCC+HCP** | **Cast** | **1941** | **1200** | [36] |
| HfTaTiZr | BCC | Cast | 1941 | 1400 | [36] |
| **$Hf_{11}Ti_{44}V_{22}Zr_{22}$** | **BCC** | **Cast** | **1941** | **2300** | [37] |
| MoNbTaTi | BCC | Anneal | 1673 | 400 | [16] |
| MoNbTaTi | BCC | AM | 1941 | 400 | [38] |
| $Mo_7Nb_{36}Ta_{18}Ti_{36}$ | BCC | AM | 1941 | 300 | [39] |
| MoNbTaV | BCC | Cast | 2183 | 1200 | [40] |
| MoNbTaW | BCC | Cast | 2750 | 500 | [13] |
| MoNbTaW | BCC | AM | 2750 | 500 | [41] |
| MoNbTaW | BCC | AM | 2750 | 500 | [42] |
| MoNbTiV | BCC | Cast | 1941 | 1200 | [43] |
| MoNbTiZr | BCC | Cast | 1941 | 1100 | [44] |

| Composition | Expt. Reported Phases | Processing Method | $T_{proc}$ | $T_{misc}$ | Ref. |
|---|---|---|---|---|---|
| MoTaTiV | BCC | Cast | 1941 | 1100 | [45] |
| **MoTaTiW** | **BCC+HCP** | **AM** | **1941** | **700** | [46] |
| NbTaTiV | BCC | Cast | 1941 | 1600 | [47] |
| NbTaTiW | BCC | Anneal | 1673 | 500 | [16] |
| **$Nb_{12}Ta_{25}Ti_{25}Zr_{37}$** | **BCC+Sec.** | **Cast** | **1941** | **1400** | [14] |
| **$Nb_{37}Ta_{25}Ti_{25}Zr_{12}$** | **BCC+Sec.** | **Cast** | **1941** | **800** | [14] |
| NbTaTiZr | BCC | Cast | 1941 | 1100 | [14] |
| $Nb_{30}Ta_{10}Ti_{30}Zr_{30}$ | BCC | Cast | 1941 | 900 | [14] |
| NbTaTiZr | BCC | AM | 1941 | 1100 | [48] |
| NbTaTiZr | BCC | AM | 1941 | 1100 | [49] |
| $Nb_{10}Ta_{10}Ti_{40}Zr_{40}$ | BCC | AM | 1941 | 1200 | [50] |
| NbTaVW | BCC | Cast | 2183 | 1500 | [47] |
| $NbTiV_2Zr$ | 2Bcc + Sec | Anneal | 1473 | 2700 | [32] |
| NbTiVZr | BCC + Sec | Anneal | 1473 | 2100 | [32] |
| CrHfNbTiZr | BCC + Laves | Anneal | 1173 | 3700 | [19] |
| CrMoNbTiW | BCC + Laves | Powder | 1273 | 2400 | [51] |
| CrNbTiVZr | BCC + Sec. | Anneal | 1473 | 3200 | [32] |
| $Cr_{24}Ta_{24}Ti_4V_{24}W_{24}$ | BCC + Laves | Powder | 1773 | 2100 | [33] |
| $Cr_{23}Ta_{23}Ti_7V_{23}W_{23}$ | BCC + Laves | Powder | 1773 | 2200 | [33] |
| HfMoNbTaTi | BCC | Cast | 1941 | 700 | [52] |
| $Hf_5Mo_{15}Nb_{35}Ta_{25}Ti_{20}$ | BCC + Sec. | AM | 1941 | 3900 | [53] |
| HfMoNbTaZr | BCC | Cast | 2128 | 1200 | [54] |
| $Hf_{25}Mo_{12}Nb_{25}Ti_{25}V_{12}$ | BCC | Cast | 1941 | 1600 | [55] |
| $Hf_{13}Mo_6Nb_{13}Ti_{41}V_{27}$ | BCC + HCP | AM | 1941 | 3900 | [56] |
| **$Hf_{12}Mo_{12}Nb_{25}Ti_{25}Zr_{25}$** | **BCC + Sec.** | **Cast** | **1941** | **1000** | [57] |
| $Hf_{12}Mo_{12}Nb_{25}Ti_{25}Zr_{25}$ | BCC | Cast | 1941 | 1000 | [58] |
| HfMoNbTiZr | BCC | Anneal | 1373 | 900 | [59] |
| HfMoTaTiZr | BCC | Cast | 1941 | 1300 | [54] |
| **HfNbTaTiZr** | **BCC + HCP** | **Anneal** | **1073** | **1000** | [17] |
| HfNbTaTiZr | BCC | Anneal | 1473 | 1000 | [18] |
| HfNbTaTiZr | 2Bcc+HCP | Anneal | 773 | 1000 | [17] |
| HfNbTaTiZr | BCC | Cast | 1941 | 1000 | [52] |
| $Hf_8Nb_{31}Ta_{31}Ti_{18}Zr_{13}$ | BCC | Cast | 1941 | 900 | [60] |
| $Hf_{12}Nb_{12}Ta_{12}Ti_{38}Zr_{25}$ | BCC | Cast | 1941 | 900 | [60] |
| $Hf_{15}Nb_{20}Ta_{10}Ti_{30}Zr_{25}$ | BCC | Anneal | 1270 | 900 | [60] |
| $Hf_{28}Nb_5Ta_5Ti_{35}Zr_{28}$ | BCC | Anneal | 1270 | 900 | [60] |
| HfNbTaTiZr | BCC | AM | 1941 | 1000 | [61] |
| $Hf_5Nb_{30}Ta_5Ti_{30}Zr_{30}$ | BCC | AM | 1941 | 700 | [62] |
| **HfNbTiVZr** | **BCC** | **Anneal** | **873** | **2200** | [19] |
| **$Hf_{11}Nb_5Ti_{42}V_{21}Zr_{21}$** | **BCC** | **Cast** | **1941** | **2100** | [37] |
| $Hf_{10}Nb_{10}Ti_{40}V_{20}Zr_{20}$ | BCC | Cast | 1941 | 1900 | [37] |
| $Hf_{10}Nb_{14}Ti_{38}V_{19}Zr_{19}$ | BCC | Cast | 1941 | 1900 | [37] |
| $Hf_9Nb_{18}Ti_{36}V_{18}Zr_{18}$ | BCC | Cast | 1941 | 1600 | [37] |

| Composition | Expt. Reported Phases | Processing Method | $T_{proc}$ | $T_{misc}$ | Ref. |
|---|---|---|---|---|---|
| HfNbTiVZr | BCC + Laves | Anneal | 1073 | 2200 | [19] |
| MoNbTaTiV | BCC | Cast | 1941 | 1200 | [63] |
| MoNbTaTiW | BCC | Anneal | 1673 | 500 | [30] |
| $Mo_{24}Nb_{24}Ta_{24}Ti_6W_{24}$ | BCC | Cast | 1941 | 400 | [64] |
| $Mo_{22}Nb_{22}Ta_{22}Ti_{11}W_{22}$ | BCC | Cast | 1941 | 500 | [64] |
| $Mo_{21}Nb_{21}Ta_{21}Ti_{16}W_{21}$ | BCC | Cast | 1941 | 500 | [64] |
| MoNbTaTiW | BCC | AM | 1941 | 500 | [65] |
| MoNbTaTiW | BCC | AM | 1941 | 500 | [66] |
| **MoNbTaTiZr** | **BCC + BCC** | **Anneal** | **1273** | **1100** | [67] |
| **MoNbTaTiZr** | **BCC + BCC** | **AM** | **1941** | **1200** | [68] |
| MoNbTaVW | BCC | Cast | 2183 | 1300 | [13] |
| MoNbTaVW | BCC | AM | 2183 | 1300 | [69] |
| MoNbTaVW | BCC | AM | 2183 | 1300 | [70] |
| $Mo_{24}Nb_{24}Ti_{24}V_6Zr_{24}$ | BCC | Anneal | 1300 | 1300 | [71] |
| $Mo_{14}Nb_{14}Ti_{14}V_{43}Zr_{14}$ | BCC | Cast | 1941 | 1900 | [44] |
| $Mo_{17}Nb_{17}Ti_{17}V_{33}Zr_{17}$ | BCC | Cast | 1941 | 1900 | [44] |
| $Mo_{18}Nb_{18}Ti_{18}V_{27}Zr_{18}$ | BCC | Cast | 1941 | 1900 | [44] |
| $Mo_{21}Nb_{21}Ti_{21}V_{16}Zr_{21}$ | BCC | Cast | 1941 | 1700 | [44] |
| $Mo_{22}Nb_{22}Ti_{22}V_{11}Zr_{22}$ | BCC | Cast | 1941 | 1500 | [44] |
| MoNbTiVZr | BCC + Laves | Anneal | 1473 | 1800 | [31] |
| MoNbTiVZr | BCC | Cast | 1941 | 1800 | [71] |
| NbTaTiVW | BCC | Cast | 1941 | 1300 | [47] |
| NbTaTiVZr | BCC + BCC | Anneal | 1473 | 1700 | [31] |
| $Cr_{12}Mo_{20}Nb_{42}Ta_1Ti_{13}V_{12}$ | BCC + Sec. | Powder | 1473 | 2500 | [72] |
| $Cr_{20}Mo_{10}Nb_{20}Ta_{10}Ti_{20}Zr_{20}$ | 2Bcc + Laves | Anneal | 1473 | 3000 | [31] |
| $Cr_{20}Mo_{10}Nb_{20}Ta_{10}Ti_{20}Zr_{20}$ | 2Bcc + Sec. | Cast | 1941 | 3000 | [73] |
| CrMoNbTaVW | BCC + Laves + Sec. | Powder | 1773 | 2400 | [74] |
| $Hf_{19}Mo_5Nb_{19}Ta_{19}Ti_{19}Zr_{19}$ | BCC | Cast | 1941 | 1100 | [52] |
| $Hf_{18}Mo_9Nb_{18}Ta_{18}Ti_{18}Zr_{18}$ | BCC | Cast | 1941 | 1100 | [52] |
| $Hf_{17}Mo_{13}Nb_{17}Ta_{17}Ti_{17}Zr_{17}$ | BCC | Cast | 1941 | 1100 | [52] |
| HfMoNbTaTiZr | BCC | Cast | 1941 | 1100 | [52] |
| **HfNbTaTiWZr** | **BCC + Laves** | **Cast** | **1941** | **1400** | [75] |
| MoNbTaTiVW | BCC | Cast | 1941 | 1000 | [31] |
| **HfMoNbTaTiWZr** | **BCC + Laves** | **Cast** | **1941** | **1400** | [75] |

## Predictive accuracy of the model on the experimental dataset

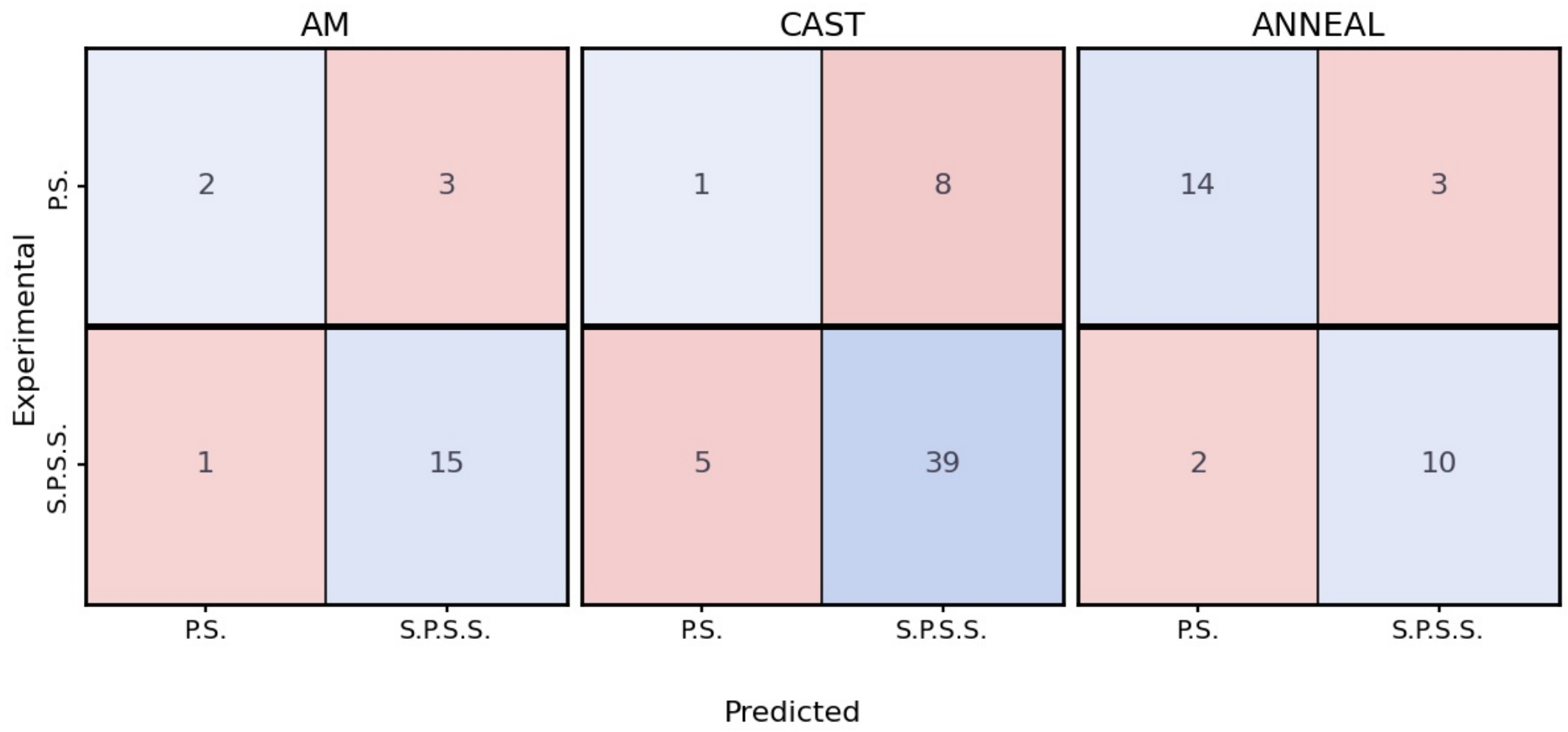

**Fig. 7. Confusion matrices comparing experimental (Expt.) and predicted (Pred.) phase distributions for phase-segregated (P.S.) and single-phase solid solution (SPSS) samples.** The left, middle, and right panels represent different processing conditions: additive manufacturing (AM), as-cast (CAST), and heat-processed (ANNEAL), respectively. While the full dataset includes six sintered samples (discussed in the main text), they are excluded from this figure. Blue diagonal and red off-diagonal terms indicate correct and incorrect predictions, respectively.

The experimental dataset comprises of three primary processing categories: As-cast, Additive Manufacturing (AM), and Annealed. As-cast samples, 53 of 109, are quenched, but depending on each case, the quenching rate might not be fast enough and therefore secondary phases can be precipitate at lower temperatures. In contrast, thermally annealed samples—with holding times ranging from one hour to several days—more closely approach equilibrium, though hot isostatic pressing (HIP) may stabilize metastable phases due to specific stress-strain conditions. AM samples, primarily synthesized via directed energy deposition (DED) or powder-bed fusion (PBF), are characterized by rapid, non-equilibrium solidification [9]. Notably, the dataset includes six sintered samples, all of which exhibit phase segregation as seen from Table 1. We also observe that identical compositions can yield different microstructures depending on the processing route; for example, CrMoTaTi, CrNbTaTi, CrNbTiW, HfMoNbTiZr, and HfNbTiVZr exhibit varying stability (SPSS vs. phase segregated) across different reports, as seen from Table 1.

Of the 29 heat-treated samples, 12 (41%) formed a SPSS. In contrast, 44 of the 53 as-cast samples (83%) remained as SPSS, a trend consistent with the sluggish diffusion effects often attributed to high-entropy systems. Consequently, if an SPSS is predicted to be stable at elevated temperatures, it is highly probable that as-cast samples will retain this single-phase structure experimentally. AM samples similarly show a high propensity for SPSS formation. Overall, the model correctly classifies 87 of 109 compositions, yielding an 80% accuracy. The primary discrepancies occur in as-cast samples, where the model predicts SPSS for 8 samples that are experimentally phase segregated; conversely, five are predicted segregated but observed as SPSS. Additionally, five annealed and four AM samples were misclassified, as seen from Fig. 7. For as-cast and AM samples, $T_{proc}$ is the lowest melting temperature of the constituent elements, a conservative proxy for the solidus temperature, so this comparison tests retention of the high-temperature solid solution. In contrast, the 29 annealed samples, whose $T_{proc}$ is an experimentally reported temperature and therefore independent of this proxy, are classified with 83% accuracy (24 of 29). Fig. 7 illustrates the breakdown of the model predictions against experimental observations, categorized by processing method. The following section discusses these discrepancies and evaluates the model's predictive limits. It should be noted that eight systems, four quaternary and four quinary have been discussed in the main text and are excluded from discussion here. All predictions in this appendix, including the phase fractions at the experimental processing temperatures, can be reproduced using the Raptor web interface: https://raptor.engr.wustl.edu and the accompanying TDB files.

## Misclassifications by the model

We analyze the misclassified systems and compositions reported to undergo phase segregation, starting with the annealed samples.

***Phase segregation misclassifications.*** Equimolar CrMoNbTi was experimentally observed to remain stable even after annealing at 1673 K [30], whereas our model predicts segregation into a 75% BCC

+ 25% secondary BCC mixture at the same temperature. We attribute this discrepancy to an overestimation of the miscibility gap in the Cr–Mo binary from 1200 K to 2000 K, as discussed in the preceding article.

Another interesting example of an HCP phase segregation is the HfNbTaZr equimolar alloy. Maiti et al. [34] annealed a sample at 2073 K for up to 8 days. Until 4 days of annealing, an HCP phase was not observed, and the sample showed only a single-phase BCC solution. However, a minor HCP phase rich in Hf-Zr was formed after 4 days, which we fail to predict at such a high temperature. None of the HCP phases are supposed to be stable at such high temperatures, because the pure elements themselves transition to the BCC phase at 1500 K. Only at 800 K, do we predict a BCC + HCP phase segregation, and a single-phase BCC solution at temperatures above 1100 K. Thus, unequal cooling across the sample could have resulted in the HCP phase segregating out.

Finally, Todai et al. [67] synthesized a MoNbTaTiZr quinary and annealed the sample at 1273 K and observed a dual BCC phase segregation. We predict a single-phase solution above 1100 K, and a BCC + HCP phase segregation at 1000 K. The elastic energies of the different phases might penalize the formation of an HCP phase and hence lead to our misclassification.

Thus, from the above discussion, it should be noted that out of the 5 misclassifications of the heat processed samples, 3 samples have miscibility temperatures close to the processing temperatures and hence are incorrectly predicted.

***Additive manufacturing misclassifications.*** Now we move on to the four incorrectly predicted AM samples. Equimolar CrMoNbV sample was reported to be a SPSS [29], but we predict the miscibility temperature to be 3000 K, far above its melting temperature. We attribute the misclassification to the overestimation of miscibility gap of Cr-Mo binary. A Zr-lean HfNbTiZr sample is predicted to have a miscible temperature of ~700 K, and hence show a stable SPSS, however a secondary precipitate is experimentally seen to form [3]. Laser deposited samples in the same alloy systems near the equimolar predicts a stable BCC solution [35]. Thus, it is possible that a secondary phase segregatedout at a lower temperature just in one sample. Equimolar MoNbTaTiZr [68] and MoTaTiW [46] synthesized by the powder bed fusion method was seen to exhibit BCC + BCC and BCC + HCP phase segregations while we predict their miscible temperature to be around 800 K. Similar phase segregations are predicted by the model at temperatures lower than 600 K. In the case of MoTaTiW [46], the TiW binary is predicted to be a solid solution at a much lower temperature, and hence is the reason for the misclassification.

***As-cast misclassifications.*** Finally, we discuss the as-cast samples, with thirteen misclassifications. In the case of the $Hf_{11}Ti_{44}V_{22}Zr_{22}$ sample [39][37], we predict the alloy's miscible temperature to be above its processing temperature proxy of 1941 K (the melting temperature of Ti), while it experimentally exhibits a BCC solid solution. The experimentally observed BCC solid solution may represent a kinetically trapped state. Notably, the $ZrV_2$ Laves phase is predicted to be stable in this system and should segregate even at elevated temperatures. Four other as-cast samples, CrMoTaTi, CrNbTaTi, CrNbTiW, and $Hf_{11}Nb_5Ti_{42}V_{21}Zr_{21}$, are likewise predicted to segregate because their $T_{misc}$ (2100–2500 K) exceeds the $T_{proc}$ of 1941 K set by Ti, whereas they are experimentally observed as BCC solid solutions, consistent with a kinetically trapped high-temperature state; the CrMoTaTi, CrNbTaTi, and CrNbTiW samples annealed at 1673 K do show secondary phases.

Next, we discuss the as-cast systems that we predict to be SPSS but is experimentally found to be phase segregated. It is possible that the different cooling conditions or annealing times in these systems could have caused secondary phase segregation at lower temperatures. We first assess the temperatures at which the model predicts the experimentally observed phase segregation occurs for these samples.

Huang et al. [36] observed a BCC + HCP phase segregation to occur as Ta was decreased from the equimolar HfTaTiZr alloy in as-cast samples. At temperatures lower than 1200 K we too predict a two-phase BCC phase segregation.

Wang et al. [75] synthesized the equimolar senary and septenary HfNbTaTiWZr and HfMoNbTaTiWZr systems that showed a dual BCC phase segregation in the as-cast structure. CALPHAD predictions reported in the paper also backed up their experimental report. We however predict a single phase to be stable in both these alloys at 1400 K. At 1200 K, we predict a dual BCC phase segregation. Therefore, we predict the same phase segregations albeit at lower temperatures.

A similar conclusion can be drawn for NbTaTiZr samples synthesized by Zyka et al. [14] by arc melting, which show segregation into multiple BCC phases; we predict this segregation at 1000 K for the Zr-rich sample, but only at or below about 700 K for the Nb-rich sample (Fig. 10(a,b)).

Taken together, these comparisons highlight that we predict phase segregation for these as-cast samples at lower temperatures, and the different quenching rates do not kinetically trap the SPSS for these samples. Resolving such discrepancies will likely require further calculations of diffusion properties, which play a critical role in distinguishing kinetically trapped phases from true equilibrium states. A deeper analysis into why some as-cast samples are trapped in a solid solution, while others do not, is beyond the scope of this paper but is an avenue for future work.

## Phase segregation predictions by the model

Next, we briefly discuss the correctly predicted phase segregated systems. We mainly focus on annealed samples, as the phase segregation could compare well to near equilibrium conditions.

***Phase segregations predicted for annealed samples.*** We start with equimolar samples of CrMoTaTi, CrNbTaTi by Coury et al. [16], homogenized at 1673 K. The CrMoTaTi and CrNbTaTi structure displayed a two-phase microstructure with a BCC + C15 Laves precipitate enriched in Cr. At 1600 K we predict a phase segregation for CrMoTaTi containing 32% phase fraction of the $TaCr_2$ Laves phase with 68% of a BCC SPSS. Similarly, for CrNbTaTi we predict two phases; 31% $TaCr_2$ Laves phase and 67% of a BCC SPSS.

Next, we analyze the phase segregations in equimolar CrNbTiZr and CrNbTiVZr alloys reported by Senkov et al. [32]. Both samples were annealed at 1473 K. For both alloys they report a Laves intermetallic + BCC SPSS microstructure. For CrNbTiZr, we predict a phase segregation of 7% Cr-rich BCC + 28% $TiCr_2$ Laves Phase + 65% NbTiZr BCC SPSS. The Cr-rich SPSS constituting <10% of phases may not form experimentally, depending on the nucleation and interface energies of the phase. For the CrNbTiVZr system we predict a phase segregation of nearly equal phase fractions of two BCC phases, CrV rich and NbTiZr rich. We however do not predict a Laves phase segregation.

Senkov et al. [32] also synthesized equimolar NbTiVZr and $NbTiV_2Zr$ alloys, annealing them at 1473K. For the equimolar alloy we predict an 80% BCC phase + 20% of a $ZrV_2$ Laves phase at the same temperature. The microstructure reported by Senkov et al. [32] also reported a dominant BCC phase with V enriched precipitates likely pointing to the Laves intermetallic. On the other hand, the $NbTiV_2Zr$ experimentally shows a phase segregation of 3 BCC phases. We predict 2 BCC phases enriched and depleted in V + 25% of the $ZrV_2$ Laves phase. The reported microstructure is complex but observation of two BCC phases with V-rich and V-lean compositions match well with our results.

Senkov et al. [31] predicted a BCC SPSS + Laves phase segregation at 1273K for the senary $CrMo_{0.5}NbTa_{0.5}TiZr$. We predict a 70% BCC + 22% $TaCr_2$ + 8% $TiCr_2$ Laves phase segregation, aligning accurately with the experimental reports. They also reported that different databases from the PANDAT software also predicted a BCC + Laves phase segregation, agreeing with our predictions.

***Phase segregations by powder method synthesis***. Next, we move on to discuss alloys synthesized by powder methods like sintering. We correctly predict phase segregation to occur in all these samples, but differences remain in the segregated phases. Waseem et al. [33] analyzed the effect of Ti addition to the CrTaVW system at a final temperature of 1773 K through a multi-step sintering process. The CrTaVW equimolar quaternary alloy is reported to have a three phase Laves phase segregation: 43%, 14%, and 43% (by volume) of W-, V-, and Cr-rich phases. We predict a dual BCC phase segregation: 56% and 44% of TaW- and CrV-rich phases. Since Cr-V and Ta-W have highly negative binary mixing enthalpies such a phase segregation is understandable in our prediction. The addition of 4% and 7% Ti to the equimolar though results in a dual BCC phase segregation, consistent with our results.

Similarly, Yan et al. [51] prepared a CrMoNbTiW equimolar quinary using a spark plasma sintering process at 1673 K. The alloy showed a phase segregation consisting of W-rich solid solution matric with a small amount of Ti-rich solid solution and Laves phase. However, at 1673 K, we predict a dual BCC phase segregation of 18% Cr-rich and 82% MoNbTiW-rich phase. For the above-mentioned sintered samples, we misclassify the nature of phase segregations for the samples. Sintering is a non-equilibrium process, involving a number of mechanisms [76], hence this could be one reason for our predictions to not align with the experimental reports.

In summary, we effectively predict the miscibility of solid solutions in multinary phase spaces. The nature of phase segregations that occur in these systems can vary widely with processing methods along with annealing temperatures and times, and we also predict the phase segregations correctly for many samples as discussed in the preceding section. As-cast samples show metastable single-phase solutions, while further annealing leads to precipitation of intermetallics like in CrMoTaTi, CrMoNbTi etc., or lower-order BCC phases as in HfNbTiVZr based on the nucleation and interphase energies. Our simplified approach can predict the phase segregations in most of the annealed samples and the stabilization of a solid solution in many as-cast and AM samples, highlighting the method's efficacy.

## Appendix B. Percentage miscible regions of quaternary and quinary RCCAs

### Table 2. Quinary alloy systems exhibiting high-range PMR (65-100%)

| Alloy | Percentage Miscible Region (%) | | |
|---|---|---|---|
| | 500 K | 1000 K | 1500 K |
| Mo-Nb-Ta-V-W | 6 | 25 | 100 |
| Mo-Nb-Ti-V-W | 3 | 55 | 100 |
| Mo-Nb-Ta-Ti-W | 77 | 100 | 100 |
| Mo-Ta-Ti-V-W | 8 | 72 | 100 |
| **Hf-Nb-Ta-Ti-Zr** | 6 | 56 | 100 |
| Hf-Mo-Nb-Ta-Ti | 17 | 86 | 99 |
| Hf-Mo-Nb-Ti-Zr | 0 | 68 | 96 |
| **Hf-Mo-Nb-Ta-Zr** | 0 | 14 | 94 |
| Nb-Ta-Ti-V-W | 0 | 15 | 94 |
| **Mo-Nb-Ta-Ti-Zr** | 0 | 35 | 94 |
| Mo-Nb-Ta-Ti-V | 1 | 21 | 93 |
| Hf-Mo-Ta-Ti-Zr | 0 | 20 | 87 |
| Hf-Nb-Ta-Ti-W | 0 | 46 | 79 |
| Hf-Mo-Ta-Ti-V | 3 | 14 | 75 |
| Hf-Mo-Ta-Ti-W | 1 | 20 | 68 |
| Hf-Mo-Nb-Ti-W | 0 | 18 | 66 |

### Table 3. Quinary alloy systems exhibiting mid-range PMR (25-65%)

| Alloy | Percentage Miscible Region (%) | | |
|---|---|---|---|
| | 500 K | 1000 K | 1500 K |
| Hf-Mo-Nb-Ti-V | 1 | 7 | 57 |
| Hf-Mo-Nb-Ta-V | 0 | 7 | 56 |
| Mo-Nb-Ti-W-Zr | 0 | 14 | 56 |
| Nb-Ta-Ti-W-Zr | 0 | 11 | 55 |
| Hf-Ta-Ti-V-W | 0 | 4 | 54 |
| Cr-Mo-Ti-V-W | 1 | 6 | 49 |
| Hf-Nb-Ti-W-Zr | 0 | 11 | 46 |
| Hf-Mo-Ti-V-W | 0 | 8 | 45 |
| Hf-Nb-Ta-Ti-V | 0 | 0 | 45 |
| Hf-Mo-Ta-V-W | 0 | 3 | 42 |
| Mo-Nb-Ti-V-Zr | 0 | 0 | 37 |

| Alloy | Percentage Miscible Region (%) | | |
|---|---|---|---|
| | 500 K | 1000 K | 1500 K |
| Hf-Nb-Ti-V-W | 0 | 6 | 34 |
| Mo-Ta-Ti-V-Zr | 0 | 0 | 34 |
| Mo-Nb-Ta-V-Zr | 0 | 0 | 32 |
| Mo-Nb-Ta-W-Zr | 0 | 1 | 31 |
| Hf-Mo-Nb-V-W | 0 | 1 | 30 |
| Hf-Nb-Ta-W-Zr | 0 | 3 | 28 |
| Mo-Ta-Ti-W-Zr | 0 | 3 | 28 |
| Hf-Mo-Ti-W-Zr | 0 | 3 | 28 |
| Hf-Nb-Ta-V-W | 0 | 0 | 27 |

## Table 4. Quaternary alloy systems exhibiting high-range PMR (65-100%)

| Alloy | Percentage Miscible Region (%) | | |
|---|---|---|---|
| | 500K | 1000K | 1500K |
| Nb-Ta-Ti-W | 56 | 100 | 100 |
| Mo-Ta-Ti-W | 21 | 100 | 100 |
| Mo-Nb-Ti-W | 75 | 100 | 100 |
| Mo-Ti-V-W | 11 | 95 | 100 |
| Mo-Nb-Ta-W | 80 | 100 | 100 |
| Mo-Nb-Ta-Ti | 72 | 100 | 100 |
| Hf-Nb-Ti-Zr | 21 | 86 | 100 |
| Hf-Nb-Ta-Ti | 24 | 87 | 100 |
| Mo-Nb-Ti-V | 0 | 45 | 99 |
| Hf-Nb-Ta-Zr | 1 | 33 | 98 |
| Nb-Ta-Ti-Zr | 4 | 53 | 96 |
| Nb-Ti-V-W | 4 | 25 | 94 |
| Mo-Nb-V-W | 9 | 34 | 94 |
| Mo-Ta-V-W | 4 | 47 | 93 |
| Hf-Mo-Ta-Ti | 4 | 54 | 91 |
| Mo-Nb-Ti-Zr | 2 | 48 | 91 |
| Hf-Mo-Nb-Ti | 32 | 71 | 89 |
| Hf-Mo-Nb-Ta | 4 | 53 | 89 |
| Hf-Ta-Ti-Zr | 0 | 15 | 87 |
| Ta-Ti-V-W | 0 | 31 | 86 |
| Hf-Mo-Ti-Zr | 6 | 53 | 85 |
| Mo-Ta-Ti-V | 0 | 38 | 84 |
| Hf-Mo-Nb-Zr | 0 | 28 | 82 |
| Cr-Ti-V-W | 0 | 14 | 73 |
| Mo-Nb-Ta-Zr | 0 | 8 | 73 |
| Mo-Nb-Ta-V | 2 | 16 | 72 |

| Alloy | Percentage Miscible Region (%) | | |
|---|---|---|---|
| | 500K | 1000K | 1500K |
| Nb-Ta-V-W | 0 | 4 | 69 |
| Mo-Ta-Ti-Zr | 0 | 14 | 66 |
| Hf-Ta-Ti-W | 0 | 28 | 65 |

## Table 5. Quaternary alloy systems exhibiting mid-range PMR (25-65%)

| Alloy | Percentage Miscible Region (%) | | |
|---|---|---|---|
| | 500K | 1000K | 1500K |
| Hf-Nb-Ti-W | 4 | 31 | 57 |
| Cr-Mo-Ti-V | 0 | 11 | 57 |
| Hf-Nb-Ta-W | 0 | 25 | 55 |
| Cr-Mo-V-W | 1 | 18 | 54 |
| Hf-Mo-Ti-V | 1 | 7 | 49 |
| Nb-Ti-W-Zr | 0 | 24 | 49 |
| Hf-Mo-Ta-V | 0 | 0 | 45 |
| Hf-Mo-Ti-W | 0 | 11 | 42 |
| Hf-Mo-Ta-Zr | 0 | 2 | 39 |
| Hf-Ta-Ti-V | 0 | 0 | 36 |
| Hf-Mo-Nb-W | 0 | 5 | 34 |
| Hf-Mo-Ta-W | 0 | 5 | 32 |
| Mo-Ti-W-Zr | 0 | 7 | 28 |
| Hf-Mo-Nb-V | 0 | 1 | 28 |
| Hf-Nb-Ti-V | 0 | 1 | 28 |
| Cr-Ta-Ti-V | 0 | 0 | 27 |
| Hf-Ti-W-Zr | 0 | 8 | 27 |
| Hf-Ti-V-W | 0 | 1 | 27 |
| Nb-Ta-W-Zr | 0 | 5 | 26 |

## Appendix C. Propagation of miscibility for mid-range quinary PMR systems

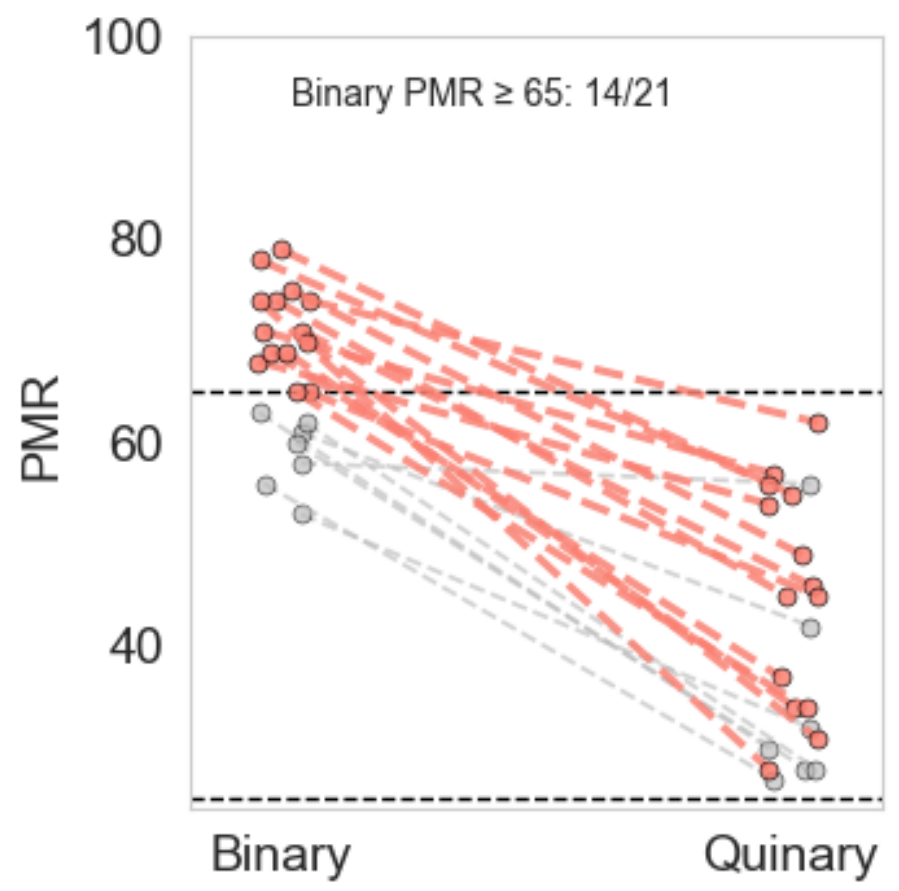

**Fig. 8. Evolution of phase-miscibility ratio from binary to mid-range quinary PMR systems**. Each dashed line connects the average PMR of a binary subspace to the corresponding quinary space, with slight horizontal jitter added for visual clarity. The horizontal dashed lines separate high and mid-range PMR regimes. Highlighted trajectories indicate binary systems with PMR above 65, showing how highly miscible binaries lead to mid-range PMR systems.

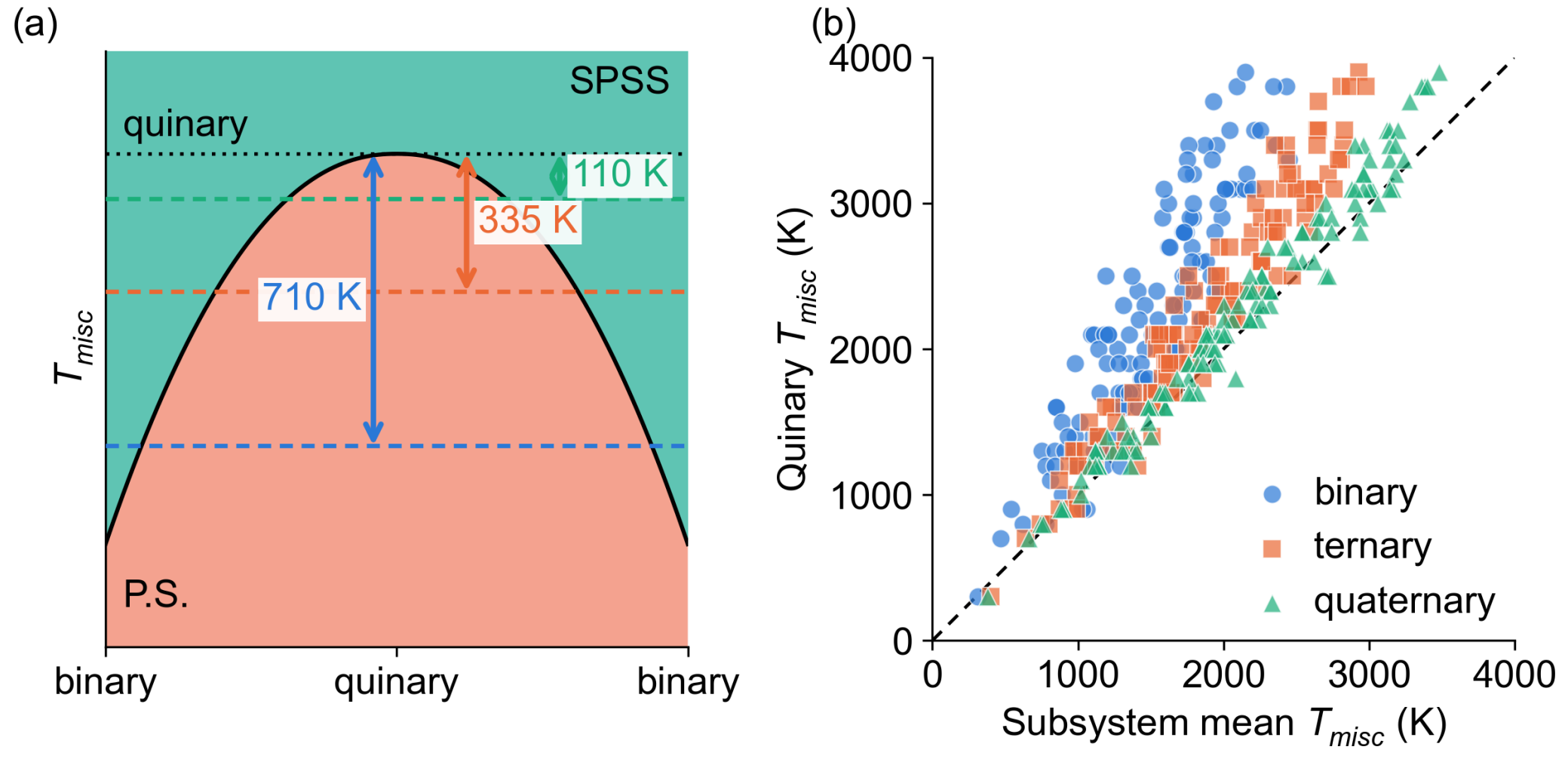

**Fig. 9. Increase of the miscibility temperature from lower-order subsystems to the equimolar quinary.** (a) Schematic of the miscibility boundary (black curve) along a composition path through the quinary center; the dashed lines mark the mean $T_{misc}$ of the binary, ternary, and quaternary subsystems, and the arrows give the shift to the quinary $T_{misc}$ (medians of 710, 335, and 110 K). (b) For each of the 126 equimolar quinary systems, the quinary $T_{misc}$ (vertical axis) is plotted against the mean $T_{misc}$ of its 10 binary (blue circles), 10 ternary (orange squares), and 5 quaternary (green triangles) subsystems. The dashed black line marks zero shift, i.e., equality of the quinary and subsystem-average $T_{misc}$, so the vertical distance of a point above it is the shift of the quinary relative to its subsystem average.

## Appendix D. Predicted phase fractions at the temperatures discussed in the text

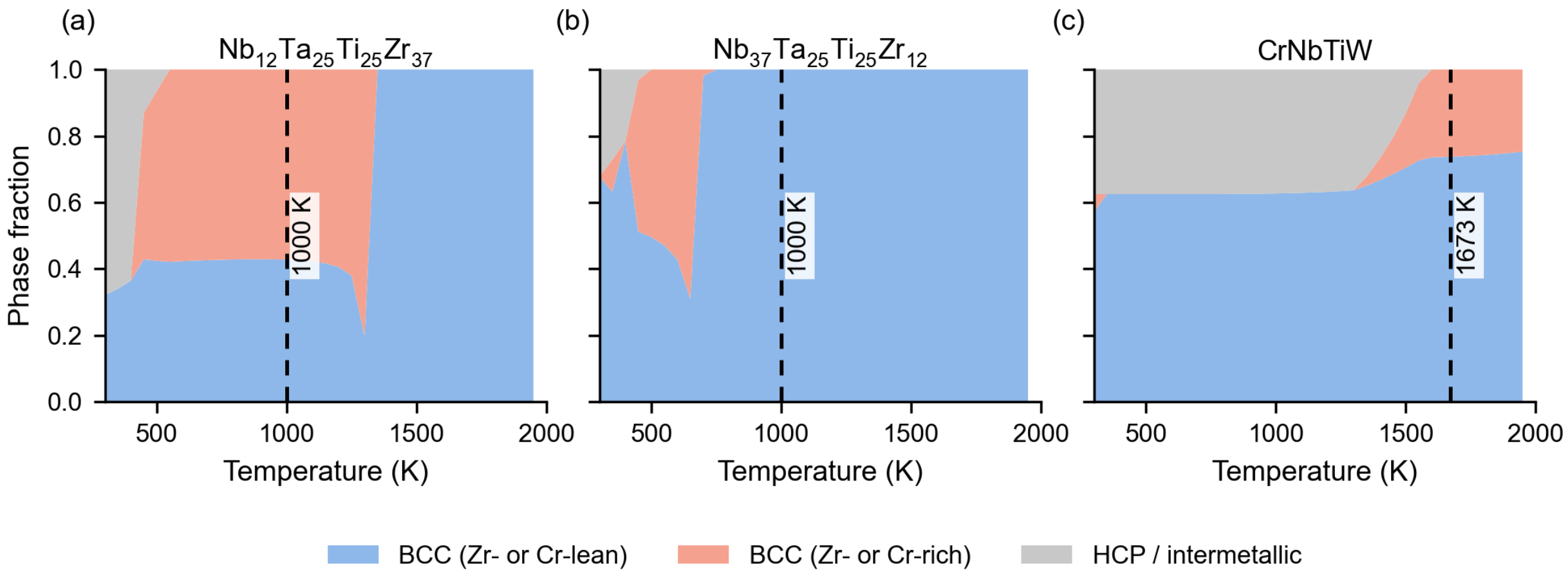

**Fig. 10. Predicted equilibrium phase fractions as a function of temperature for (a) $Nb_{12}Ta_{25}Ti_{25}Zr_{37}$ and (b) $Nb_{37}Ta_{25}Ti_{25}Zr_{12}$, the Zr-rich and Nb-rich as-cast samples of Zyka et al., and (c) equimolar CrNbTiW.** Where two BCC phases coexist they are ordered by Zr (a, b) or Cr (c) content, and the grey region denotes HCP and intermetallic phases. Dashed lines mark the temperatures discussed in the text: 1000 K (a, b) and 1673 K (c).

## References

[1] D.B. Miracle, O.N. Senkov, A critical review of high entropy alloys and related concepts, Acta Materialia 122 (2017) 448-511.
[2] O.N. Senkov, D.B. Miracle, K.J. Chaput, J.-P. Couzinie, Development and exploration of refractory high entropy alloys—A review, J Mater Res 33(19) (2018) 3092-3128.
[3] S. Gou, M. Gao, Y. Shi, S. Li, Y. Fang, X. Chen, H. Chen, W. Yin, J. Liu, Z. Lei, Additive manufacturing of ductile refractory high-entropy alloys via phase engineering, Acta Materialia 248 (2023) 118781.
[4] Y. Zhang, Z. Li, J. Xie, X. Zhao, H. Chen, Y. Wang, J.-F. Nie, Strain-induced fully coherent triphase nanoarchitecture in refractory high-entropy alloys, Science 392(6804) (2026) 1308-1312.
[5] D.H. Cook, P. Kumar, M.I. Payne, C.H. Belcher, P. Borges, W. Wang, F. Walsh, Z. Li, A. Devaraj, M. Zhang, Kink bands promote exceptional fracture resistance in a NbTaTiHf refractory medium-entropy alloy, Science 384(6692) (2024) 178-184.
[6] B. Sun, M. Fedorov, J. Wrobel, D. Nguyen-Manh, C. Hatler, M. Vigil, D.J. Thoma, O. El-Atwani, E. Martinez, Thermodynamic assessment of the quaternary WTaCrV refractory high entropy alloy as a means to guide experimental approaches, Acta Materialia 299 (2025).
[7] O. El Atwani, H.T. Vo, M.A. Tunes, C. Lee, A. Alvarado, N. Krienke, J.D. Poplawsky, A.A. Kohnert, J. Gigax, W.Y. Chen, M. Li, Y.Q. Wang, J.S. Wróbel, D. Nguyen-Manh, J.K.S. Baldwin, O.U. Tukac, E. Aydogan, S. Fensin, E. Martinez, A quinary WTaCrVHf nanocrystalline refractory high-entropy alloy withholding extreme irradiation environments, Nature Communications 14(1) (2023) 2516.
[8] S. Gorsse, M.H. Nguyen, O.N. Senkov, D.B. Miracle, Database on the mechanical properties of high entropy alloys and complex concentrated alloys, Data in Brief 21 (2018) 2664-2678.
[9] W. Zhang, A. Chabok, B.J. Kooi, Y. Pei, Additive manufactured high entropy alloys: A review of the microstructure and properties, Materials & Design 220 (2022).
[10] A. Mohammadnejad, M. Sun, N.A. Yarlapati, M. Hasanabadi, E. Sadeghi, P. Karimi, Y. Zou, E. Toyserkani, Additive manufacturing of refractory high-entropy alloys: A critical review of fundamentals and advances, Materials & Design 260 (2025) 115261.
[11] W. Chen, A. Hilhorst, G. Bokas, S. Gorsse, P.J. Jacques, G. Hautier, A map of single-phase high-entropy alloys, Nat Commun 14(1) (2023) 2856.
[12] J. Cavin, P. Omprakash, A. Couet, R. Mishra, SymPlex Plots for Visualizing Properties in High-Dimensional Alloy Spaces, arXiv, 2025.
[13] O.N. Senkov, G.B. Wilks, J.M. Scott, D.B. Miracle, Mechanical properties of Nb25Mo25Ta25W25 and V20Nb20Mo20Ta20W20 refractory high entropy alloys, Intermetallics 19(5) (2011) 698-706.
[14] J. Zyka, J. Malek, J. Vesely, F. Lukac, J. Cizek, J. Kuriplach, O. Melikhova, Microstructure and Room Temperature Mechanical Properties of Different 3 and 4 Element Medium Entropy Alloys from HfNbTaTiZr System, Entropy (Basel) 21(2) (2019) 114.
[15] A.C.R.M. Pravan Omprakash; Nicholas Crnkovich; John Cavin; Nathan Curtis, First-principles-based Prediction of Phase Fields, 10.5281/zenodo.18866462, 2026.
[16] F.G. Coury, M. Kaufman, A.J. Clarke, Solid-solution strengthening in refractory high entropy alloys, Acta Materialia 175 (2019) 66-81.
[17] B. Schuh, B. V√∂lker, J. Todt, N. Schell, L. Perri√®re, J. Li, J.P. Couzini√©, A. Hohenwarter, Thermodynamic instability of a nanocrystalline, single-phase TiZrNbHfTa alloy and its impact on the mechanical properties, Acta Materialia 142 (2018) 201-212.
[18] O.N. Senkov, S.L. Semiatin, Microstructure and properties of a refractory high-entropy alloy after cold working, Journal of Alloys and Compounds 649 (2015) 1110-1123.
[19] â. Fazakas, V. Zadorozhnyy, L.K. Varga, A. Inoue, D.V. Louzguine-Luzgin, F. Tian, L. Vitos, Experimental and theoretical study of Ti20Zr20Hf20Nb20X20 (X ¬† = ¬† V or Cr) refractory high-entropy alloys, International Journal of Refractory Metals and Hard Materials 47 (2014) 131-138.
[20] J.L. Li, Z. Li, Q. Wang, C. Dong, P.K. Liaw, Phase-field simulation of coherent BCC/B2 microstructures in high entropy alloys, Acta Materialia 197 (2020) 10-19.
[21] Binary Alloy Phase Diagrams, (2016).

[22] M. Li, Z.H. Zhang, A.S. Thind, G.D. Ren, R. Mishra, K. Flores, Microstructure and properties of NbVZr refractory complex concentrated alloys, Acta Materialia 213 (2021).
[23] T. Butler, K. Chaput, J. Dietrich, O. Senkov, High temperature oxidation behaviors of equimolar NbTiZrV and NbTiZrCr refractory complex concentrated alloys (RCCAs), Journal of Alloys and Compounds 729 (2017) 1004-1019.
[24] F. Liu, P. Liaw, Y. Zhang, Recent Progress with BCC-Structured High-Entropy Alloys, Metals 12(3) (2022).
[25] G.B. Bokas, W. Chen, A. Hilhorst, P.J. Jacques, S. Gorsse, G. Hautier, Unveiling the thermodynamic driving forces for high entropy alloys formation through big data ab initio analysis, Scripta Materialia 202 (2021).
[26] Z.H. Zhang, M. Li, J. Cavin, K. Flores, R. Mishra, A Fast and Robust Method for Predicting the Phase Stability of Refractory Complex Concentrated Alloys using Pairwise Mixing Enthalpy, Acta Materialia 241 (2022) 118389.
[27] M.C. Troparevsky, J.R. Morris, P.R. Kent, A.R. Lupini, G.M. Stocks, Criteria for predicting the formation of single-phase high-entropy alloys, Physical Review X 5(1) (2015) 011041.
[28] H. Luan, L. Huang, J. Kang, B. Luo, X. Yang, J. Li, Z. Han, J. Si, Y. Shao, J. Lu, K.-F. Yao, Spinodal decomposition and the pseudo-binary decomposition in high-entropy alloys, Acta Materialia 248 (2023) 118775.
[29] Z. He, B. Ouyang, Charting the electronic structure for discovering low-cost intermetallic catalysts, Journal of Materials Chemistry A 12(33) (2024) 21987-21996.
[30] F.G. Coury, T. Butler, K. Chaput, A. Saville, J. Copley, J. Foltz, P. Mason, K. Clarke, M. Kaufman, A. Clarke, Phase equilibria, mechanical properties and design of quaternary refractory high entropy alloys, Materials & Design 155 (2018) 244-256.
[31] O.N. Senkov, S. Gorsse, D.B. Miracle, High temperature strength of refractory complex concentrated alloys, Acta Materialia 175 (2019) 394-405.
[32] O.N. Senkov, S.V. Senkova, D.B. Miracle, C. Woodward, Mechanical properties of low-density, refractory multi-principal element alloys of the Cr–Nb–Ti–V–Zr system, Materials Science and Engineering: A 565 (2013) 51-62.
[33] O.A. Waseem, J. Lee, H.M. Lee, H.J. Ryu, The effect of Ti on the sintering and mechanical properties of refractory high-entropy alloy TixWTaVCr fabricated via spark plasma¬†sintering for fusion plasma-facing materials, Materials Chemistry and Physics 210 (2018) 87-94.
[34] S. Maiti, W. Steurer, Structural-disorder and its effect on mechanical properties in single-phase TaNbHfZr high-entropy alloy, Acta Materialia 106 (2016) 87-97.
[35] Y.-M. Hu, X.-D. Liu, N.-N. Guo, L. Wang, Y.-Q. Su, J.-J. Guo, Microstructure and mechanical properties of NbZrTi and NbHfZrTi alloys, Rare Met. 38(9) (2019) 840-847.
[36] H. Huang, Y. Wu, J. He, H. Wang, X. Liu, K. An, W. Wu, Z. Lu, Phase-Transformation Ductilization of Brittle High-Entropy Alloys via Metastability Engineering, Advanced Materials 29(30) (2017) 1701678.
[37] D.-X. Qiao, H. Jiang, W.-N. Jiao, Y.-P. Lu, Z.-Q. Cao, T.-J. Li, A Novel Series of Refractory High-Entropy Alloys Ti2ZrHf0.5VNbx with High Specific Yield Strength and Good Ductility, Acta Metall. Sin. (Engl. Lett.) 32(8) (2019) 925-931.
[38] H. Zhang, Y. Zhao, J. Cai, S. Ji, J. Geng, X. Sun, D. Li, High-strength NbMoTaX refractory high-entropy alloy with low stacking fault energy eutectic phase via laser additive manufacturing, Materials & Design 201 (2021) 109462.
[39] F. Wang, T. Yuan, R. Li, S. Lin, P. Niu, V. Cristino, Effect of Mo on the morphology, microstructure and mechanical properties of NbTa0. 5TiMox refractory high entropy alloy fabricated by laser powder bed fusion using elemental mixed powders, International Journal of Refractory Metals and Hard Materials 111 (2023) 106107.
[40] H. Yao, J.-W. Qiao, M.C. Gao, J.A. Hawk, S.-G. Ma, H. Zhou, MoNbTaV Medium-Entropy Alloy, Entropy 18(5) (2016) 189.
[41] H. Zhang, Y. Zhao, S. Huang, S. Zhu, F. Wang, D. Li, Manufacturing and analysis of high-performance refractory high-entropy alloy via selective laser melting (SLM), Materials 12(5) (2019) 720.

[42] H. Zhang, W. Xu, Y. Xu, Z. Lu, D. Li, The thermal-mechanical behavior of WTaMoNb high-entropy alloy via selective laser melting (SLM): experiment and simulation, The International Journal of Advanced Manufacturing Technology 96(1) (2018) 461-474.
[43] S.Y. Chen, X. Yang, K.A. Dahmen, P.K. Liaw, Y. Zhang, Microstructures and crackling noise of AlxNbTiMoV high entropy alloys, Entropy 16(2) (2014) 870-884.
[44] Y. Zhang, X. Yang, P.K. Liaw, Alloy Design and Properties Optimization of High-Entropy Alloys, JOM 64(7) (2012) 830-838.
[45] D.X. Qiao, H. Jiang, X.X. Chang, Y.P. Lu, T.J. Li, Microstructure and mechanical properties of VTaTiMoAlx refractory high entropy alloys, Materials science forum, Trans Tech Publ, 2017, pp. 638-642.
[46] C. Liu, K. Zhu, W. Ding, Y. Liu, G. Chen, X. Qu, Additive manufacturing of WMoTaTi refractory high-entropy alloy by employing fluidised powders, Powder Metallurgy 65(5) (2022) 413-425.
[47] H. Yao, J. Qiao, M. Gao, J. Hawk, S. Ma, H. Zhou, Y. Zhang, NbTaV-(Ti, W) refractory high-entropy alloys: Experiments and modeling, Materials Science and Engineering: A 674 (2016) 203-211.
[48] S. Mooraj, G. Kim, X. Fan, S. Samuha, Y. Xie, T. Li, J.S. Tiley, Y. Chen, D. Yu, K. An, Additive manufacturing of defect-free TiZrNbTa refractory high-entropy alloy with enhanced elastic isotropy via in-situ alloying of elemental powders, Communications Materials 5(1) (2024) 14.
[49] D. Preisler, T. Krajňák, M. Janeček, J. Kozlík, J. Stráský, M. Brázda, J. Džugan, Directed energy deposition of bulk Nb-Ta-Ti-Zr refractory complex concentrated alloy, Materials Letters 337 (2023) 133980.
[50] D.K. Jobes, Y. Liu, L. Lopez, S. Oh, A. Bucsek, D. Rubio-Ejchel, C. Tandoc, Y.-J. Hu, J.V. Gordon, Probing rapid solidification pathways in refractory complex concentrated alloys via multimodal synchrotron X-ray imaging and melt pool-scale simulation, J Mater Res 40(1) (2025) 81-97.
[51] J. Yan, K. Li, Y. Wang, J. Qiu, Microstructure and Mechanical Properties of WMoNbCrTi HEAs Sintered from the Powders Milled for Different Durations, JOM 71(8) (2019) 2489-2497.
[52] C.-C. Juan, K.-K. Tseng, W.-L. Hsu, M.-H. Tsai, C.-W. Tsai, C.-M. Lin, S.-K. Chen, S.-J. Lin, J.-W. Yeh, Solution strengthening of ductile refractory HfMoxNbTaTiZr high-entropy alloys, Materials Letters 175 (2016) 284-287.
[53] A. Ozalp, C. Okuyucu, B. Koç, O. El-Atwani, E. Aydogan, Development and directed energy deposition of high strength Hf5Mo15Nb35Ta25Ti20 refractory high entropy alloys, Materials Characterization 209 (2024) 113679.
[54] K.-K. Tseng, C.-C. Juan, S. Tso, H.-C. Chen, C.-W. Tsai, J.-W. Yeh, Effects of Mo, Nb, Ta, Ti, and Zr on Mechanical Properties of Equiatomic Hf-Mo-Nb-Ta-Ti-Zr Alloys, Entropy 21(1) (2019) 15.
[55] Y. Liu, Y. Zhang, H. Zhang, N. Wang, X. Chen, H. Zhang, Y. Li, Microstructure and mechanical properties of refractory HfMo0. 5NbTiV0. 5Six high-entropy composites, Journal of Alloys and Compounds 694 (2017) 869-876.
[56] D. Cui, B. Guo, Z. Yang, X. Liu, Z. Wang, J. Li, J. Wang, F. He, Unraveling microstructure and mechanical response of an additively manufactured refractory TiVHfNbMo high-entropy alloy, Additive Manufacturing 84 (2024) 104126.
[57] N. Guo, L. Wang, L. Luo, X. Li, R. Chen, Y. Su, J. Guo, H. Fu, Microstructure and mechanical properties of refractory high entropy (Mo0. 5NbHf0. 5ZrTi) BCC/M5Si3 in-situ compound, Journal of alloys and compounds 660 (2016) 197-203.
[58] N. Guo, L. Wang, L. Luo, X. Li, R. Chen, Y. Su, J. Guo, H. Fu, Microstructure and mechanical properties of in-situ MC-carbide particulates-reinforced refractory high-entropy Mo0. 5NbHf0. 5ZrTi matrix alloy composite, Intermetallics 69 (2016) 74-77.
[59] N. Guo, L. Wang, L. Luo, X. Li, Y. Su, J. Guo, H. Fu, Microstructure and mechanical properties of refractory MoNbHfZrTi high-entropy alloy, Materials & Design 81 (2015) 87-94.
[60] A.V. Podolskiy, E.D. Tabachnikova, V.V. Voloschuk, V.F. Gorban, N.A. Krapivka, S.A. Firstov, Mechanical properties and thermally activated plasticity of the Ti30Zr25Hf15Nb20Ta10 high entropy alloy at temperatures 4.2,Äì350 ¬† K, Materials Science and Engineering: A 710 (2018) 136-141.

[61] H. Dobbelstein, E.P. George, E.L. Gurevich, A. Kostka, A. Ostendorf, G. Laplanche, Laser metal deposition of refractory high-entropy alloys for high-throughput synthesis and structure-property characterization, International Journal of Extreme Manufacturing 3(1) (2021) 015201.
[62] Z. Liang, Y. Wen, L. Sun, X. Li, Y. Zhang, High temperature phase transformation and mechanical properties of Hf0. 25NbTa0. 25TiZr high-entropy alloys formed by laser melting deposition (LMD), Materials Science and Engineering: A 913 (2024) 147021.
[63] H. Yao, J. Qiao, J. Hawk, H. Zhou, M. Chen, M. Gao, Mechanical properties of refractory high-entropy alloys: Experiments and modeling, Journal of Alloys and Compounds 696 (2017) 1139-1150.
[64] Z. Han, H. Luan, X. Liu, N. Chen, X. Li, Y. Shao, K. Yao, Microstructures and mechanical properties of TixNbMoTaW refractory high-entropy alloys, Materials Science and Engineering: A 712 (2018) 380-385.
[65] J. Liu, J. Li, X. Du, Y. Tong, R. Wang, D. He, Z. Cai, H. Wang, Microstructure and mechanical properties of wire arc additively manufactured MoNbTaWTi high entropy alloys, Materials 14(16) (2021) 4512.
[66] B. Xiao, W. Jia, H. Tang, J. Wang, L. Zhou, Microstructure and mechanical properties of WMoTaNbTi refractory high-entropy alloys fabricated by selective electron beam melting, Journal of Materials Science & Technology 108 (2022) 54-63.
[67] M. Todai, T. Nagase, T. Hori, A. Matsugaki, A. Sekita, T. Nakano, Novel TiNbTaZrMo high-entropy alloys for metallic biomaterials, Scripta Materialia 129 (2017) 65-68.
[68] G. Carlucci, J. Fiocchi, E. Ferrario, C. Biffi, R. Casati, Improvement of LPBF processability of equiatomic MoNbTaTiZr RHEA powder by in-situ precipitation of boride particles, Materials Letters 371 (2024) 136944.
[69] P. Gu, T. Qi, L. Chen, T. Ge, X. Ren, Manufacturing and analysis of VNbMoTaW refractory high-entropy alloy fabricated by selective laser melting, International Journal of Refractory Metals and Hard Materials 105 (2022) 105834.
[70] L. Chen, Z. Yang, L. Lu, X. Zhang, X. Ren, Effect of TiC on the high-temperature oxidation behavior of WMoTaNbV refractory high entropy alloy fabricated by selective laser melting, International Journal of Refractory Metals and Hard Materials 110 (2023) 106027.
[71] Y.D. Wu, Y.H. Cai, X.H. Chen, T. Wang, J.J. Si, L. Wang, Y.D. Wang, X.D. Hui, Phase composition and solid solution strengthening effect in TiZrNbMoV high-entropy alloys, Materials & Design 83 (2015) 651-660.
[72] B. Kang, T. Kong, A. Raza, H.J. Ryu, S.H. Hong, Fabrication, microstructure and mechanical property of a novel Nb-rich refractory high-entropy alloy strengthened by in-situ formation of dispersoids, International Journal of Refractory Metals and Hard Materials 81 (2019) 15-20.
[73] O. Senkov, C. Woodward, Microstructure and properties of a refractory NbCrMo0. 5Ta0. 5TiZr alloy, Materials Science and Engineering: A 529 (2011) 311-320.
[74] Y. Long, X. Liang, K. Su, H. Peng, X. Li, A fine-grained NbMoTaWVCr refractory high-entropy alloy with ultra-high strength: Microstructural evolution and mechanical properties, Journal of Alloys and Compounds 780 (2019) 607-617.
[75] M. Wang, Z. Ma, Z. Xu, X. Cheng, Microstructures and mechanical properties of HfNbTaTiZrW and HfNbTaTiZrMoW refractory high-entropy alloys, Journal of Alloys and Compounds 803 (2019) 778-785.
[76] S. Momjian, J.C. Mauro, C.A. Randall, Non-equilibrium thermodynamics and kinetics of the sintering process: A relaxation model describing cold sintering, Journal of the American Ceramic Society n/a(n/a) e70180.